# Towards Fully Automated Segmentation of Rat Cardiac MRI by Leveraging Deep Learning Frameworks

Daniel Fernández-Llaneza,[*,†,§] Andrea Gondová,[†, §] Harris Vince,[†] Arijit Patra,[†] Magdalena Zurek,[†] Peter Konings,[‡] Patrik Kagelid,[†] Leif Hultin[†]

[†] Clinical Pharmacology and Safety Sciences, Biopharmaceuticals R&D, AstraZeneca, Pepparedsleden 1, SE 431 83 Mölndal, Sweden

[‡] Data Sciences & Quantitative Biology, Discovery Sciences, Biopharmaceuticals R&D, AstraZeneca, Pepparedsleden 1, SE 431 83 Mölndal, Sweden

[§] DF and AG contributed equally to this work

---

[*] e-mail: daniel.fernandez1@astrazeneca.com (D. Fernández-Llaneza)

## ABSTRACT

**Background**

Automated segmentation of human cardiac magnetic resonance datasets has been steadily improving during recent years. However, these methods are cannot be directly transferred in a preclinical context due to the limited number of datasets and lower image resolution. Successful application of deep architectures 3D cardiac segmentation for rats, albeit critical for cardiac function preclinical evaluation, has to our knowledge not yet been reported.

**Methods**

We developed segmentation models that expand on the standard U-Net architecture and evaluated models separately trained for systole and diastole phases, 2MSA, and a single model trained for all phases, 1MSA. Furthermore, we calibrated model outputs using a Gaussian Process (GP)-based prior to improve phase selection.

**Results**

The resulting models approach human performance in terms of left ventricular segmentation quality and ejection fraction (EF) estimation in both 1MSA and 2MSA settings (Sørensen-Dice score 0.91 ± 0.072 and 0.93 ± 0.032, respectively). 2MSA achieved a mean absolute difference between estimated and reference EF of 3.5 ± 2.5 %, while 1MSA resulted in 4.1 ± 3.0 %. Applying GPs to 1MSA enabled automating systole and diastole phase selection.

**Conclusion**

Both segmentation approaches were statistically equivalent. Combined with a proposed cardiac phase selection strategy, our work presents an important first step towards a fully automated segmentation pipeline in the context of rat cardiac analysis.

# INTRODUCTION

Preclinical assessment of novel cardiovascular therapeutics relies on the use of rodent models. Cardiac magnetic resonance (CMR) imaging is important to this assessment due to its ability to differentiate tissue types with good contrast. Moreover, the non-invasive nature of CMR makes it well suited for a longitudinal follow-up of the same animal during the treatment.[1,2] Changes in volumes of left ventricle (LV) and the ejection fraction (EF) over time are commonly assessed in response to treatment as a proxy estimate of rodent cardiac function.

Currently, the assessment of LV function mainly relies on the manual selection and slice-by-slice segmentation of the ventricle at end-diastole (ED) and end-systole (ES) phases of the cardiac cycle to estimate EF. As the number of animals in a preclinical study can be very large and are followed-up in multiple timepoints, this task can add hours of laborious analysis. Furthermore, the quality of the resulting segmentations depends on experts' experience and often suffers from intra- and interoperator variability.[3–5] Thus, fully automated segmentation tools to standardise and speed up the segmentation of rodent CMR datasets are highly desirable for *in vivo* cardiac efficacy studies.

The quality of semi- and fully automated methods for cardiac segmentation of difficult human CMR datasets has been improving thanks to the uptake of convolutional neural networks (CNN).[6–10] However, there is limited research applied in the preclinical cardiac 3D MRI segmentation sphere in smaller rodent datasets with only two recent examples in mice.[6–8] To our knowledge, this work presents the first statistically robust study into automating the estimation of EF in rats and expands on the well-established 3D U-Net[9,11] and its derived architectures (i.e., Attention U-Net,[12] U-Net++[13,14] and V-Net[15]).

Supported by our preliminary work and observations from other teams,[12] LV segmentation of the ES achieves consistently lower performance compared to the ED when attempted by the same model, namely, 1-model segmentation approach (1MSA). As this can lead to undesirable downstream effects on EF estimation, we trained two independent models to segment diastole and systole separately, 2-model segmentation approach (2MSA). Both 1MSA and 2MSA were benchmarked in terms of segmentation quality, robustness against noise and agreement for EF estimates compared to human-derived gold standards. Our results suggest that the presented 2MSA and 1MSA approaches are equivalent in terms of EF estimation and achieve close to human performance which highlights that both 1MSA and 2MSA methods are valid for use depending on the context.

In a similar vein to previous research in the UK Biobank data,[16] the 1MSA approach has been able to segment all cardiac phases captured during the recording of the cardiac cycle. This has allowed us to evaluate the potential of the method for full automation, that is, automated phase selection and EF estimation. Upon probing different metrics derived from LV segmentations along the cardiac cycle (LV volume, LV surface area or mid-slice area), we implemented a GP which identified systole and diastole phases and subsequently estimated EF. This was compared against a conventional fourth-polynomial degree fitting strategy and human selection of systole and diastole. In the light of this evidence, 1MSA emerged as a method with the potential to inspire similar work in other biological systems and organs in the preclinical setting.

## 1. METHODS

### 1.1. Data Acquisition

**Table 1** summarises the six segmentation datasets used in this study, consisting of short-axis 3D CMR images (multi-2D acquisition) from sham-operated rats or rats with induced myocardial infarcts (EF ranging from 41% - 67 %). Eleven to thirteen short-axis CINE time series (temporal resolution 8 ms) covering the LV and two perpendicular long-axis slices were acquired per animal. We use the term image stack to refer to the collection of 2D slices which make up a 3D CMR image of the rat LV with a resolution of 0.5x0.5x1.5 mm. All MRI scanning was performed using a 200 MHz (BioSpec 4.7T/40 USR; Bruker BioSpin, Karlsruhe, Germany) equipped with a 400 mT/m actively shielded gradient system and ParaVision software (PV6.0). A 72-mm i.d. quadrature resonator (Bruker, Ettlingen, Germany) was used. All experiments were performed in compliance with EU Directive 2010/63/EU. The protocols were approved by the local ethics committee (Göteborgs djurförsöksetisk nämnd).

**Table 1. Summary of Rat Heart Longitudinal Study Image Segmentation Datasets Used in the Experiments**

| Study name | Number of Image Stacks | Number of Timepoints* | Set |
|---|---|---|---|
| Study 1 | 126 | 3 | Train |
| Study 2 | 73 | 2 | Train |
| Study 3 | 43 | 3 | Train |
| Study 4 | 92 | 2 | Train |
| Study 5 | 116 | 3 | Test |
| Study 6 | 94 | 3 | Train(40%)/Test(60%) |

* Number of follow ups within the longitudinal study. In presented studies the timepoints were separated by a 3-month timeframe

### 1.2. Pre-processing

LV volumes were manually segmented using the software Segment version 2.2 R6289 (Medviso, http://segment.heiberg.se)[17] by an experienced operator (Op1) who was considered to be the gold standard. Additionally, the dataset Study 5 was segmented by another trained operator (Op2). The operators first cropped the image stack to ease the manual segmentation.. All training image stacks and the corresponding masks were resized by interpolation to a standard size of 12x86x98 voxels and then normalised.

For training, the whole Study 1, Study 2, Study 3 and Study 4 sets were used. Additionally, 38 image stacks from Study 6 were introduced in the training set to increase representativeness of the training set, as this dataset differed from the rest in the extent of cardiac injury of the animal subjects. After training, the model performance was evaluated against the Study 5 and Study 6 test image stacks.

### 1.3. Postprocessing

The results of the segmentation were subjected to thresholding. Subsequently, morphological operations were carried out eroding thin protrusions and closing potential holes within the LV cavity.[18]

### 1.4. Implementation Details

*Segmentation Approaches Definitions.* 1MSA models were trained both on systole and diastole phases and used for segmentation of all phases in the cardiac cycle. 2MSA models were trained separately on systole or diastole phases and used for segmentation of systole or diastole, respectively.

*Model Architectures.* A range of encoder-decoder architectures using CNNs have been implemented, namely, U-Net,[9] Attention U-Net,[12] UNet++[13] and V-Net.[15] 3D convolutions were applied and the downsampling steps were performed using a volumetric kernel of 2x2x2 voxels and a stride of 2x2x2 or 1x2x2 voxels (depending on the downsampling dimensions). Additionally, 3D padding was applied at the original image to avoid cropping the skip pathways. The padding was set to 'same' and the dilation rate was set to 1x1x1 voxels for the decoding path. Sigmoid was used as the last activation function. Batch normalisation was used after each layer and the dimensions of the tensors were passed as (N, D, H, W, C), where N is the number of sequences (mini batch), D is the number of images in a sequence, H is the height of one image in the sequence, W is the width of one image in the sequence and C is the number of channels (set to 1).

*Training.* A random sample of 10 % of the image stacks available for training was separated for validation. Each model was trained for 100 epochs and patience was set to 5 epochs on the validation set. Hyperparameter tuning was performed in a stage-wise fashion: 1) selection of loss function and number of downsampling dimensions, 2) number of blocks and layers, 3) other relevant hyperparameters. Grid search strategy was used for the first stage and random search[19] for the last two stages. The hyperparameters explored are detailed in **Table 2**.

**Table 2. Hyperparameters for Model Tuning**

| Hyperparameter | Value |
| --- | --- |
| loss function | soft Dice loss,[15] weighted soft Dice loss, binary cross-entropy and hybrid loss (binary cross-entropy + soft Dice loss) |
| downsampling dimensions | 2 (*x* and *y* axes), 3 (*x*, *y* and *z* axes) |
| number of blocks | 2, 3, 4 |
| number of layers | 2, 4, 6, 8 |
| kernel size | 2 x 2 x 2 voxels, 3 x 3 x 3 voxels, 5 x 5 x 5 voxels |
| number of filters | 8, 16, 32 |
| batch renormalisation[20] | False, True |
| activation function | Exponential Linear Unit (ELU), Gaussian Error Linear Unit (GELU), Scaled Exponential Linear Unit (SELU), swish |
| dropout | 0.0, 0.2, 0.4 |
| batch size | 8, 16, 32, 64 |
| optimiser | Adam, RMSProp |
| learning rate | 0.0001, 0.001 |
| kernel initialiser | Glorot uniform, Glorot normal, random normal |
| pooling | MaxPooling3D, AveragePooling3D |
| deconvolution | Conv3DTranspose, UpSampling3D |

The soft Dice loss is formulated as:

$$\mathcal{L}_{\mathrm{DSC}} = 1 - \frac{\sum_{i=1}^{N} t_i p_i + \epsilon}{\sum_{i=1}^{N} t_i + p_i + \epsilon} - \frac{\sum_{i=1}^{N} (1 - t_i)(1 - p_i) + \epsilon}{\sum_{i=1}^{N} 2 - t_i - p_i + \epsilon} \qquad \text{equation (1)}$$

where $T$ is the true foreground segmentation with voxel values $t_i$ and $P$ the predicted probabilistic segmentation for the mask over $i$ image elements $p_i$. The background class probability is $1 - P$. $\epsilon$ was set to 1. As a trivial extension, the weighted Dice loss function was designed to penalise misclassification at the borders of the region of interest (ROI). This objective function is given by:

$$\mathcal{L}_{\mathrm{wDSC}} = \sum_{i=1}^{N} w_{map}(t_i) \mathcal{L}_{\mathrm{DSC}} \qquad \text{equation (2)}$$

The weight map $w_{map}(t_i)$ was implemented as described in Ronneberger *et al.*[9]:

$$w_{map}(t_i) = w_c(t_i) + w_0 \, e^{\left(-\frac{(d_1(t_i) + d_2(t_i))}{2\sigma^2}\right)} \qquad \text{equation (3)}$$

setting $w_0 = 2$ and $\sigma = 1$. $w_c$ is the weight map to balance the class frequencies, $d_1$ denotes the distance to the border of the nearest cell and $d_2$ the distance to the border of the second nearest cell.

*Data Augmentation.* One or more of the following augmentation strategies were implemented at random on the original images: elastic deformations, random shifts (across the *x*-axis and the *y*-axis), rotation (-20° to 20°), scaling (contraction or expansion), blurring (Gaussian filter with $\sigma \in [0.5, 1.5]$) and gamma correction ($\gamma \sim \mathcal{N}(1, 0.1)$). The final training dataset was inflated by a factor of ten.

*Model Optimisation.* To segment the ROI, two approaches were explored: 2MSA and 1MSA. 2MSA consisted of training two separate models for each phase, whilst for 1MSA the model was trained with the whole set of images (regardless of their phase). The hyperparameters for the top-three initial best performing architectures for each approach were fine-tuned on the basis of Sørensen-Dice score (DSC).[21] Furthermore, ensembling strategies (voting and averaging) were evaluated.

### 1.5. Segmentation Approach Benchmarking

*Model Selection.* DSC distributions and convergence times were used as criteria for aiding the selection of models for 2MSA and 1MSA.

*Statistical Assessment of Segmentation Approaches.* A linear mixed model (LMM) accounting for 1MSA and 2MSA and two human readers using Maximum Likelihood with Satterthwaite degrees of freedom was fit as per equation (4).[22]

$$EF_{predicted} = EF_{actual} + \text{animal ID} + \mathbb{I}_{approach} + \mathbb{I}_{treatment} + \mathbb{I}_{reader} \quad \text{equation (4)}$$

where $\mathbb{I}$ denotes an indicator function. $\mathbb{I}_{reader}$ and animal ID are random effects and $\mathbb{I}_{approach}$ and $\mathbb{I}_{treatment}$ are the model approach and treatment arms fixed effects (see Supplementary Information). Additionally, contrasts were constructed to perform an equivalence test comparing both approaches with an equivalence margin of 2 % EF.[23]

*Noise Robustness Analysis.* Three test sets were artificially generated based on the images from the Study 5 by introducing Gaussian, Rician[24] and Rayleigh[25] distributed noise with a signal to noise ratio (SNR) of 30. The CNNs were further tested against a mixed noise scenario, where SNR was set to 20 and any of the aforementioned noise distributions was applied to the image at random. DSC was reported, alongside with EF mean absolute differences (MD) both for 1MSA and 2MSA as per equation (5):

$$MD := \mathbb{E}[|\Delta EF|] = \mathbb{E}[|EF_{estimated} - EF_{reference}|] \quad \text{equation (5)}$$

*Agreement Analysis.* The main endpoint of this study was to estimate the EF given a cardiac cycle. Volumes were estimated by calculating the sum of masks for each slice.

MD and Bland-Altman plots[26] were used to analyse agreement between reference and estimated EF for both 1MSA and 2MSA segmentation approaches.

### 1.6. Interoperator agreement

The models were benchmarked against the interoperator variability (Op1 vs Op2) with a Bland-Altman plot for the Study 5 test set to examine method interchangeability.

### 1.7. 1MSA Automation Feasibility Study

The availability of eleven to thirteen images stacks ordered along the temporal dimension allowed to represent a full cardiac cycle. Two different algorithms were devised to fit the cardiac cycle volume curve: a fourth-degree polynomial and a GP model.[27] The prior to fit the GP model to the data was a composite kernel consisting of a constant kernel multiplied by the radial-basis function (RBF) kernel (see Supplementary Information).

For automated systole and diastole phase selection, three different metrics were examined. These were the cardiac volume, slice area and surface area. The ground truth was taken as the manually selected diastole and systolic phase masks. MD and Bland-Altman plots were analysed to inform which metric would be best suited for this purpose. Once the systole and diastole phases had been selected, their respective heart volumes were used for EF estimation. The slice area was defined as the 2D slice whose area had the highest variance throughout the cardiac cycle. The surface area was the 2D surface mesh and was calculated using the Lewiner marching cubes algorithm.[28]

### 1.8. Software

We implemented the networks in Python v3.7.6, using the Keras library v2.0.8[29] and TensorFlow v1.14.0 backend.[30] Models were trained on a Tesla K80 using 75 GB of RAM. GPs were fitted using the scikit-learn v0.23.2 and the Lewiner marching cubes used scikit-image v0.17.2. R v3.6.3 was used to calculate intraclass coefficient ICC using irr v0.1.1 and the lme4 v1.1-26 and emmeans v.1.5.2-1 packages for LMM and equivalence testing.[31–33]

## 2. RESULTS

### 2.1. Model Optimisation

The Dice loss was found to be the optimal training loss reaching satisfactory scores in the range between 0.89 – 0.90. Notably, persisting numerical instability issues prevented progress in studying further loss functions (i.e., hybrid loss functions).

The hyperparameters for both the systole and diastole final models from the 2MSA are shown in **Table 5** and **Table 6**, respectively. **Table 7** shows the final models from the 1MSA. Further details are provided in Supplementary Information.

The number of downsampling dimensions, blocks and layers demonstrated to have a greater impact on DSC performance than the rest of the hyperparameters. Dropout levels > 0.2 showed to be detrimental and neither GELU nor SELU activation functions contributed to increasing performances compared to ELU. The rest of the hyperparameters (i.e., kernel initialiser, kernel size, number of filters, optimiser, etc.) caused only a marginal improvement in terms of DSC performance and all of the models ended up converging on the same values for these.

**Table 5. Final Models for Diastole Phase Model for 2MSA**

| Architecture | downsampling dimensions | blocks | layers | kernel size | number of filters | renormalisation | activation function | pooling | kernel initialiser | deconvolution | dropout | batch size | learning rate | optimiser | DSC |
|---|---|---|---|---|---|---|---|---|---|---|---|---|---|---|---|
| Attention U-Net | 3 | 4 | 6 | (3,3,3) | 16 | True | ELU | average | Glorot normal | True | 0.0 | 8 | $10^{-3}$ | Adam | 0.960 |
| U-Net++ | 2 | 3 | 8 | (3,3,3) | 16 | True | ELU | max | Glorot normal | True | 0.0 | 8 | $10^{-2}$ | Adam | 0.965 |
| V-Net | 2 | 3 | 4 | (3,3,3) | 16 | True | ELU | max | Glorot normal | True | 0.0 | 8 | $10^{-2}$ | Adam | 0.959 |

**Table 6. Final Models for Systole Phase Model for 2MSA**

| Architecture | downsampling dimensions | blocks | layers | kernel size | number of filters | renormalisation | activation function | pooling | kernel initialiser | deconvolution | dropout | batch size | learning rate | optimiser | DSC |
|---|---|---|---|---|---|---|---|---|---|---|---|---|---|---|---|
| Attention U-Net | 3 | 2 | 3 | (5,5,5) | 8 | True | swish | average | Glorot normal | False | 0.0 | 8 | $10^{-2}$ | Adam | 0.929 |
| U-Net | 2 | 5 | 2 | (3,3,3) | 16 | False | swish | max | Glorot normal | False | 0.2 | 8 | $10^{-3}$ | Adam | 0.938 |
| V-Net | 3 | 4 | 6 | (3,3,3) | 16 | True | ELU | max | Glorot normal | True | 0.0 | 8 | $10^{-2}$ | Adam | 0.934 |

**Table 7. Final Models for 1MSA**

| architecture | downsampling dimensions | blocks | layers | kernel size | number of filters | renormalisation | activation function | pooling | kernel initialiser | deconvolution | dropout | batch size | learning rate | optimiser | DSC |
|---|---|---|---|---|---|---|---|---|---|---|---|---|---|---|---|
| Attention U-Net | 3 | 2 | 3 | (3,3,3) | 16 | True | ELU | max | Glorot normal | True | 0.0 | 8 | $10^{-2}$ | Adam | 0.949 |
| U-Net | 2 | 3 | 4 | (3,3,3) | 16 | True | ELU | max | Glorot normal | True | 0.0 | 8 | $10^{-2}$ | Adam | 0.948 |
| V-Net | 3 | 3 | 2 | (3,3,3) | 16 | True | ELU | max | Glorot normal | True | 0.0 | 8 | $10^{-2}$ | Adam | 0.950 |

During the training process, our V-Net implementation converged up to two times faster (33 h for diastole 2MSA) than the rest of the CNN encoder-decoder architectures. In contrast, U-Net++ was the slowest network to train (56 h).

### 2.2. Model Selection

For the 2MSA, base models exhibit a mean DSC > 0.90 and right-skewed distributions (see **Fig. 1**). Interestingly, ensemble models provide similar statistics. Exploratory attempts carried out by bagging five Attention U-Net models for systole segmentation yielded a mean DSC > 0.92.

Indeed, there is a high degree of overlap in DSC distributions amongst the tested architectures. However, a higher degree of variability can be observed when it comes to systole-trained models. In fact, diastolic segmentations reached better segmentation quality than systolic ones. Thus, better performances and lower convergence times, prompted us to select V-Net as the final segmentation model for systolic phases and Attention U-Net for diastolic phases for the 2MSA approach.

With regards to 1MSA, performance for base models and ensembles was also comparable. U-Net was selected as the final model on grounds of parsimony (see **Fig. 2**). Over 75 % of the segmented hearts had a DSC > 0.90 regardless of the phase both for 1MSA and 2MSA (see Supplementary Information).

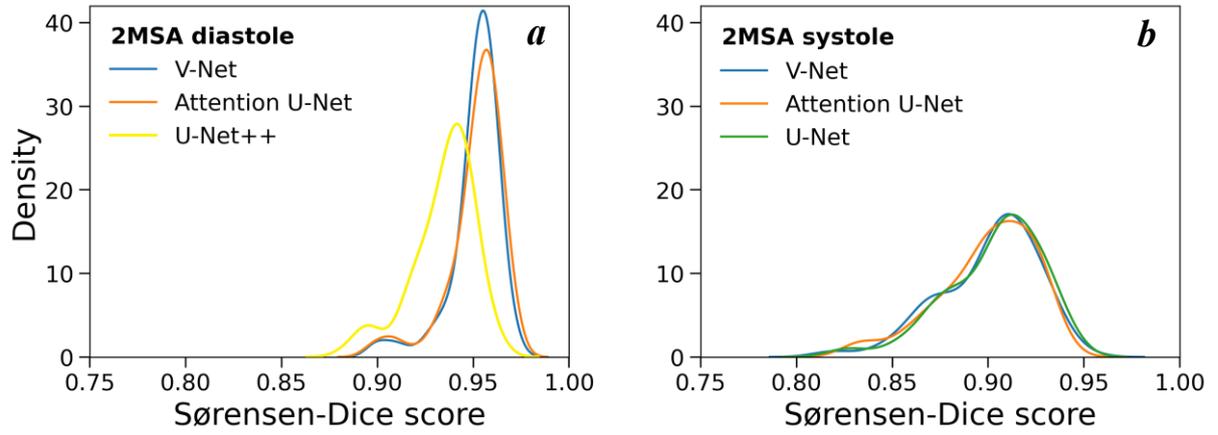

**Figure 1.** Sørensen-Dice score distribution for 2MSA models for pooled Study 5 and Study 6.
*a*: diastole models (V-Net: 0.91 ± 0.072, Attention U-Net: 0.91 ± 0.080, U-Net++: 0.90 ± 0.069)
*b*: systole models (V-Net: 0.91 ± 0.072, Attention U-Net: 0.87 ± 0.065, U-Net: 0.87 ± 0.064)

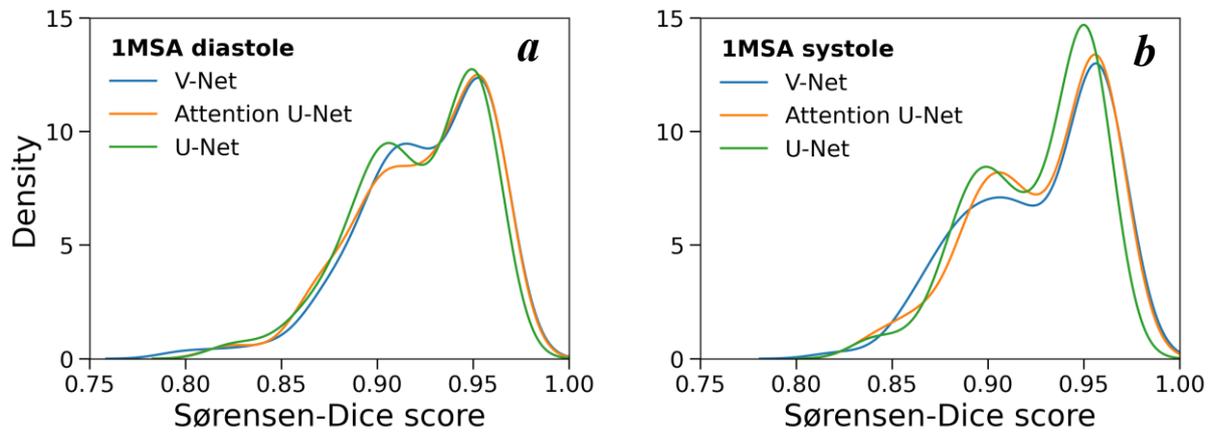

**Figure 2.** Sørensen-Dice score distribution for 1MSA models for pooled Study 5 and Study 6 where diastole and systole phases were separated manually for the sake of analysing different behaviours.
*a*: diastole segmentation (V-Net: 0.93 ± 0.032, Attention U-Net : 0.93 ± 0.032, U-Net : 0.93 ± 0.030)
*b*: systole segmentation (V-Net: 0.93 ± 0.033, Attention U-Net : 0.93 ± 0.031, U-Net : 0.92 ± 0.029)

## 2.3. Segmentation Quality Analysis

Representative examples of CNN image stack segmentations from both 1MSA and 2MSA final models are displayed in **Fig. 3**. From visual inspection of the segmented image stacks, different segmentation methods did not seem to result in any substantial differences in segmentation quality. However, systolic contours were consistently more irregular than diastolic contours. It was observed that basal and apical slices suffered from lower segmentation quality compared to midslices (see Supplementary Information).

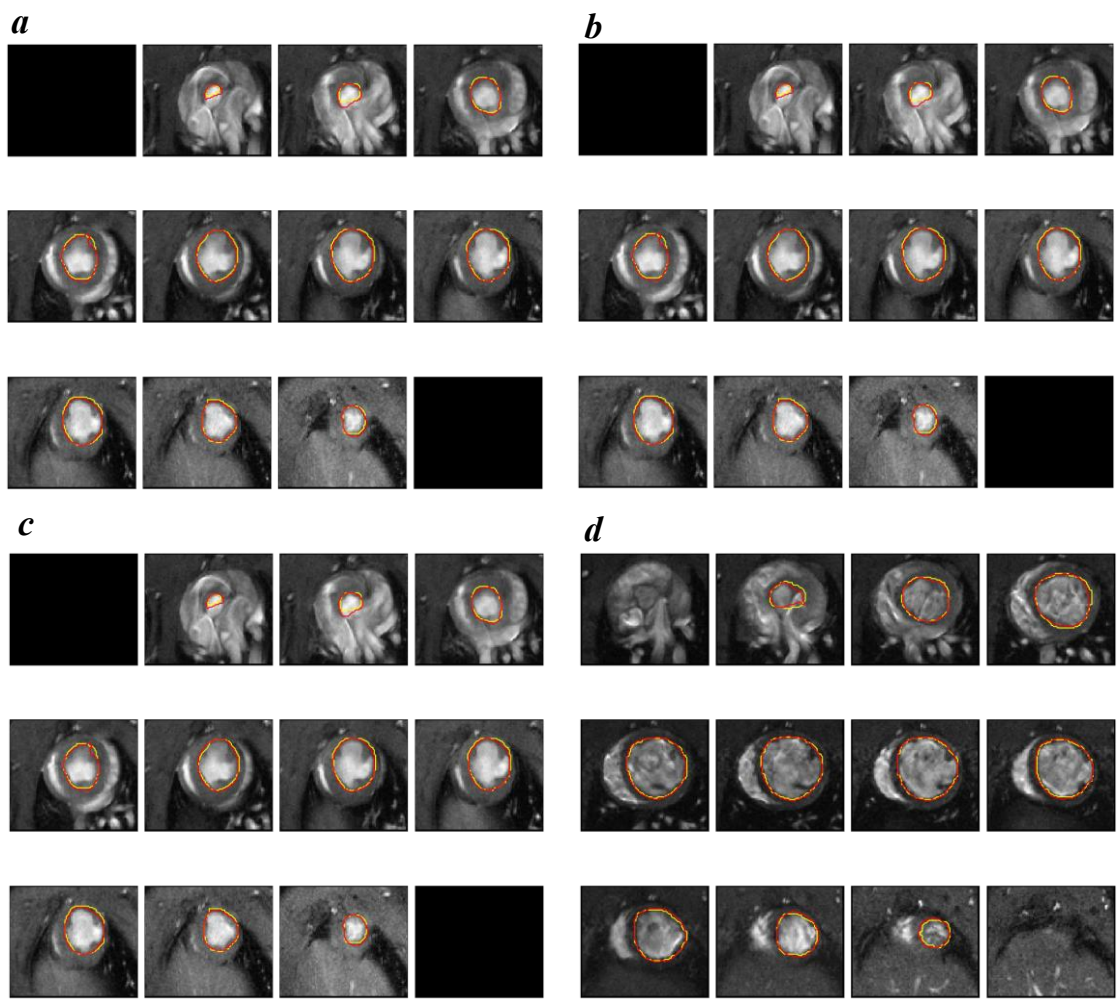

**Figure 3.** Representative examples of segmented rat hearts with U-Net . In **red** the predicted mask and in **yellow** the manual mask. *a*: systole with 1MSA. *b*: diastole with 1MSA. *c*: systole with 2MSA. *d*: diastole with 2MSA.

### 2.4. Statistical Assessment of Segmentation Approaches

The final models for both 1MSA and 2MSA were benchmarked to examine which one would be better. Consequently, a LMM was fit with the data adjusting for operator and treatment arm (see **Table 8**). The segmentation of sham-operated and myocardial infarction treatment arms was statistically significantly different ($p$-value < 0.0001) and the operator effect ($p$-value < 0.0001) too. Diagnostics confirmed that the normality assumption was tenable (see Supplementary Information). Furthermore, a locally estimated scatterplot smoothing (LOESS) fit was used to aid the interpretation of these differences (see Supplementary Information).

**Table 8. Model Table for Fixed Effects in a Linear Mixed Model**

| term | estimate | standard error | statistic | df | $p$-value |
|---|---|---|---|---|---|
| (Intercept) | 8.37 | 1.96 | 4.27 | 3.33 | 0.019 |
| human operator | 0.87 | 0.02 | 47.79 | 389.54 | < 0.0001 |
| treatment | 2.54 | 0.60 | 4.20 | 274.93 | < 0.0001 |
| model | -1.23 | 0.22 | -5.48 | 382.15 | < 0.0001 |

Additionally, we contrasted both approaches based on the LMM with an equivalence test.[23] This highlighted that the null hypothesis of the two approaches not being the same with an equivalence margin of 2% EF could be rejected ($p$-value < 0.001).

Regarding ΔEF MDs, no segmentation approach was shown to be superior, given that the absolute differences were in the same order of magnitude (see **Table 9**).

**Table 9. Ejection Fraction Mean Absolute Difference for Pooled Study 5 and Study 6** [a]

| ΔEF MD | |
|---|---|
| 2MSA | 1MSA |
| 3.5 ± 2.5 | 4.1 ± 3.0 |

[a] represented as MD ± standard deviation

## 2.5. Noise Robustness Analysis

In terms of segmentation quality, 2MSA benefits from data augmentation, whilst it does not seem to elicit major changes for 1MSA (see **Fig. 4**). The effect of data augmentation on statistics for both segmentation approaches is reported in Supplementary Information.

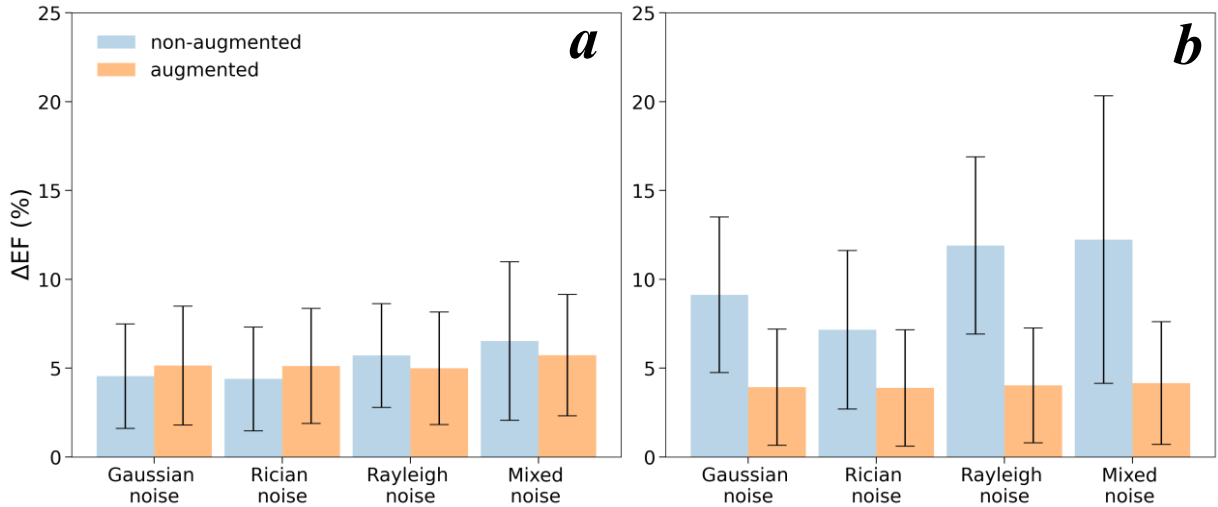

**Figure 4.** EF mean absolute difference in different noise scenarios between non-augmented (original) and augmented model for one-model segmentation approach (*left*) and two-model segmentation approach (*right*). The bars represent the standard deviation.

## 2.6. Agreement Analysis

Bland-Altman plots are displayed in **Fig. 5** for 1MSA vs Operator 1 and 2MSA vs Operator 2 approaches, respectively. **Table 10** contains the main statistics extracted from these plots.

Regardless of the segmentation approach, there was a significant negative bias ($p$-value $< 0.05$). Nevertheless, both 1MSA and 2MSA afforded comparable biases and variability. There was no clear trend upon increasing the magnitude of EF, thereby suggesting that the risk of proportional bias could be rejected.

**Table 10. Bland-Altman Plots Statistics by Segmentation Approach for Pooled Study 5 and Study 6** [a]

| segmentation approach | bias | upper bound | lower bound |
|---|---|---|---|
| 1MSA | -3.7 ± 3.5 | 3.1 | -10.5 |
| 2MSA | -2.9 ± 3.2 | 3.4 | -9.2 |

[a] represented as MD ± standard deviation

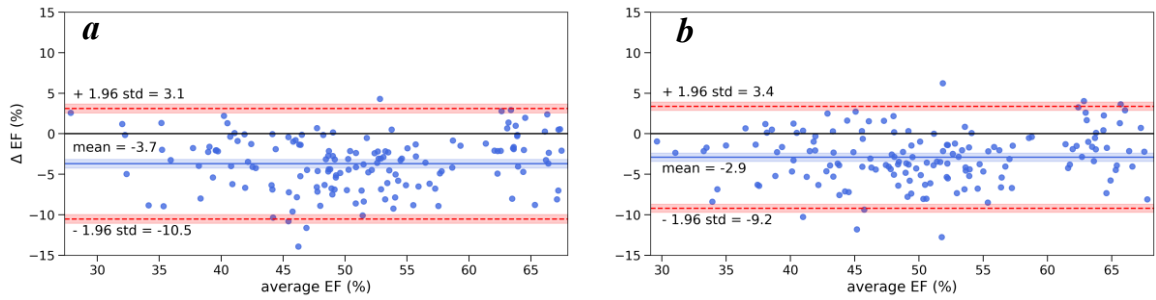

**Figure 5.** Bland-Altman plots for agreement analysis for pooled Study 5 and Study 6 using the 1MSA (bias = -3.7 ± 3.5) (***a***) and 2MSA (bias = - 2.9 ± 3.2) (***b***). The bias is marked with a solid blue line (─) and its 95% CI shaded in blue. The equality line is marked with a solid black line (─). The upper and lower 95% CI are marked with discontinuous red lines (--) and are shaded with their respective 95% CI in red.

### 2.7. Interoperator Agreement Analysis

**Fig. 6** displays the Bland-Altman analysis to assess interoperator agreement. The difference between operators is significant, as the equality line is not within the 95 % confidence interval (CI) of the mean. The average discrepancy between the two operators is -5.6 ± 3.8 %. Interoperator agreement in systole and diastole volume estimation is provided in the Supplementary Information.

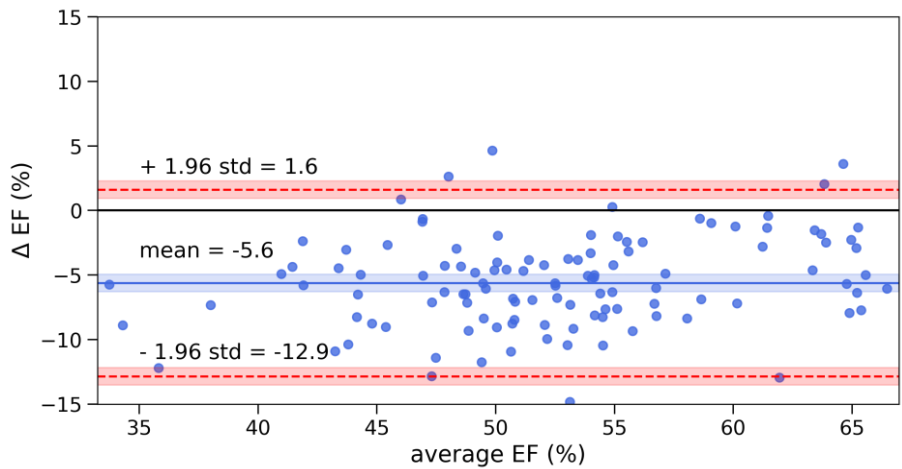

**Figure 6**. Bland-Altman plot for agreement analysis between operators for Study 5 (bias = -5.6 ± 3.8). The bias is marked with a solid blue line (─) and its 95% CI shaded in blue. The equality line is marked with a solid black line (─). The upper and lower 95% CI are marked with discontinuous red lines (--) and are shaded with their respective 95% CI.

### 2.8. 1MSA Automation Feasibility Study

The full capabilities of 1MSA could be exploited thanks to the collection of image stacks along the cardiac cycle during the MRI scanning process. **Table 11** shows the MD between the EF determined from manually segmented masks and the EF from automatic phase selection for the different selection metrics. Of note, the slice area emerged as the best metric to perform phase selection, as it had the lowest ΔEF MD both for polynomial and GP. In contrast, taking the surface area as the metric for phase selection yields the highest MDs.

**Table 11. Ejection Fraction Mean Absolute Difference Comparison for Polynomial and Gaussian Process Fitting** [a]

| fitting method | selection parameter | | |
|---|---|---|---|
| | volume | surface area | slice area |
| polynomial | 8.3 ± 4.7 | 10.5 ± 3.9 | **4.4 ± 4.1** |
| GP | 7.9 ± 4.9 | 11.3 ± 4.0 | **4.7 ± 4.3** |

[a] represented as MD ± standard deviation

Agreement both for polynomial and GP using the three selection metrics was analysed qualitatively with Bland-Altman plots (see Supplementary Information) and their statistics are summarised in **Table 12**.

In all cases, there is a significant bias ($p$-value < 0.05), given that its 95% CI does not overlap with the equality line. Interestingly, both the volume and the slice area display negative biases, while surface area overestimates the EF. The lowest bias is observed when using the slice area as the phase selection metric and the highest using volume. However, the fitting method used does not substantially impact the bias for any of the metrics.

The difference in absolute terms between the upper and the lower agreement bounds always ranges between 15 and 20. The slice area displays the broader agreement bounds, while the surface area provides the narrowest. No proportional bias is observed and the datapoints are normally distributed.

**Table 12. Bland-Altman Plots Statistics for Phase Selection Metric** [a]

| metric | method | bias | upper bound | lower bound |
|---|---|---|---|---|
| volume | polynomial | -11.5 ± 4.3 | -3.1 | -19.9 |
| | Gaussian Process | -11.2 ± 4.3 | -2.4 | -20.0 |
| surface area | polynomial | 6.7 ± 4.1 | 14.8 | -1.3 |
| | Gaussian Process | 7.4 ± 4.3 | 15.2 | -0.7 |
| **slice area** | **polynomial** | **-2.5 ± 4.9** | 7.1 | -12.1 |
| | **Gaussian Process** | **-2.5 ± 5.2** | 7.3 | -12.3 |

[a] represented as mean bias ± standard deviation

The cardiac cycle graphs were scrutinised by visual inspection to determine which method would be better placed to resolve challenging tasks. GP fitting followed the tendency marked by datapoints, whereas the polynomials adopted a generic form which oftentimes ignored finer experimental trends. An archetypical example is depicted in **Figure 7**.

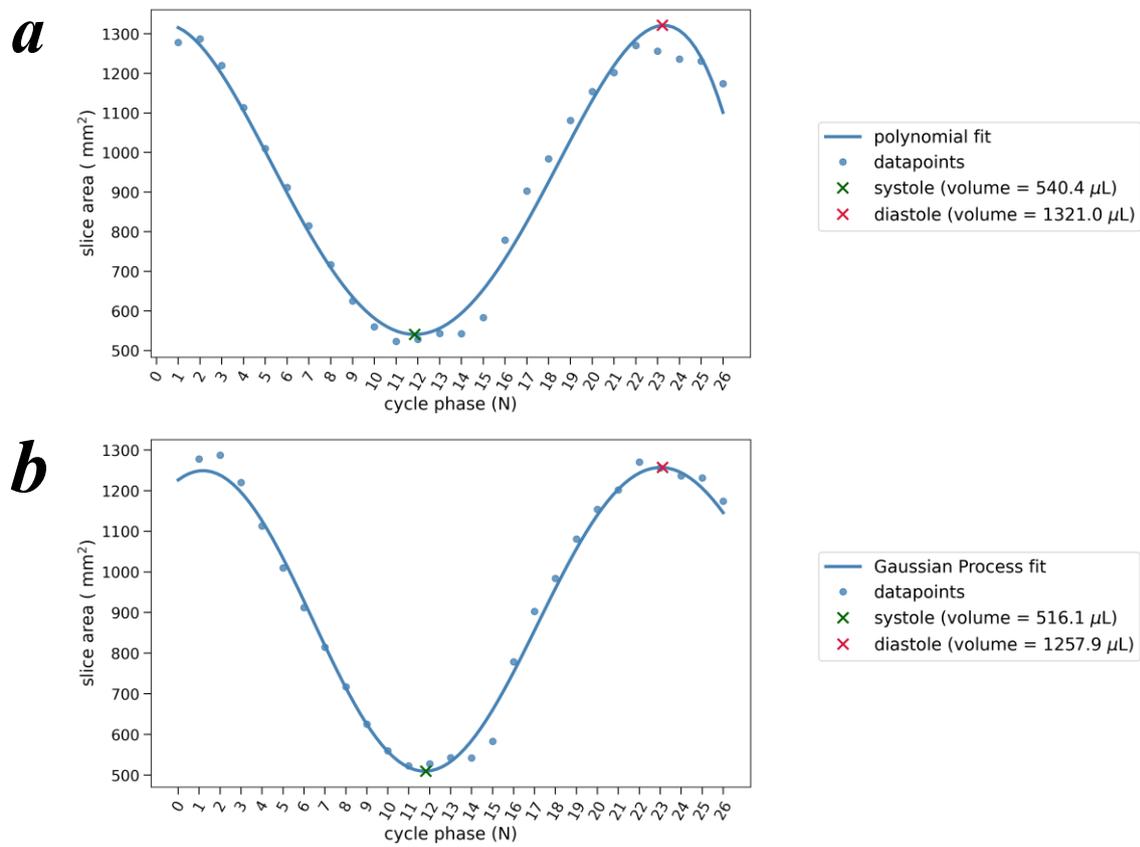

**Figure 7**. Cardiac cycle plot showing the evolution of the automatically selected slice area vs. the cycle phase. ✕ indicates the selected diastolic phase and ✕ the selected systolic phase. Their volumes are subsequently used to estimate the EF. Fourth-degree polynomial (***a***) and Gaussian Process (kernel = Constant * RBF) (***b***) fitting methods. The polynomial fit considerably overestimates the end-diastolic volume and does not select an actual datapoint for the end-systolic volume, while the GP fits the datapoints more closely and affords a better estimation of the EF.

## 3. DISCUSSION

U-Net and its derived architectures demonstrated excellent performances for rat LV cardiac segmentation. However, the extension of U-Net with attention gates,[12] deep supervision[13] or residual connections[15] did not to translate into better performances in our data. Thus, U-Nets remain as competitive and capable to successfully extract relevant information even with 2 downsampling dimensions. Therefore, simpler U-Nets render futile architectural fine-tuning and extensive skip connections. Against initial expectations,[17] ensembles have only proved non-inferiority to base models, possibly due to the correlated nature of constituent. This motivated the rationale that model selection should be based on grounds of parsimony, which in its turn reduced convergence speed. Indeed, it was noticed that complex architectures (i.e., U-Net++) required longer training times.

Interestingly, a factor that influenced the quality of the segmentation is the cardiac phase. The lower segmentation quality in systole could be caused by class imbalance and a higher variability of systole phases compared to diastolic ones (see **Fig. 8**).[34,35] This might also be the reason why basal and apical slices suffered from poorer segmentation quality. Increased interslice coherence could play a role in improving segmentation, as it was seen that using 3 downsampling dimensions did not elicit performance improvements. In effect, strategies like leveraging recurrent neural networks have been tried in similar settings[36] and it could be a potential avenue for future exploration.

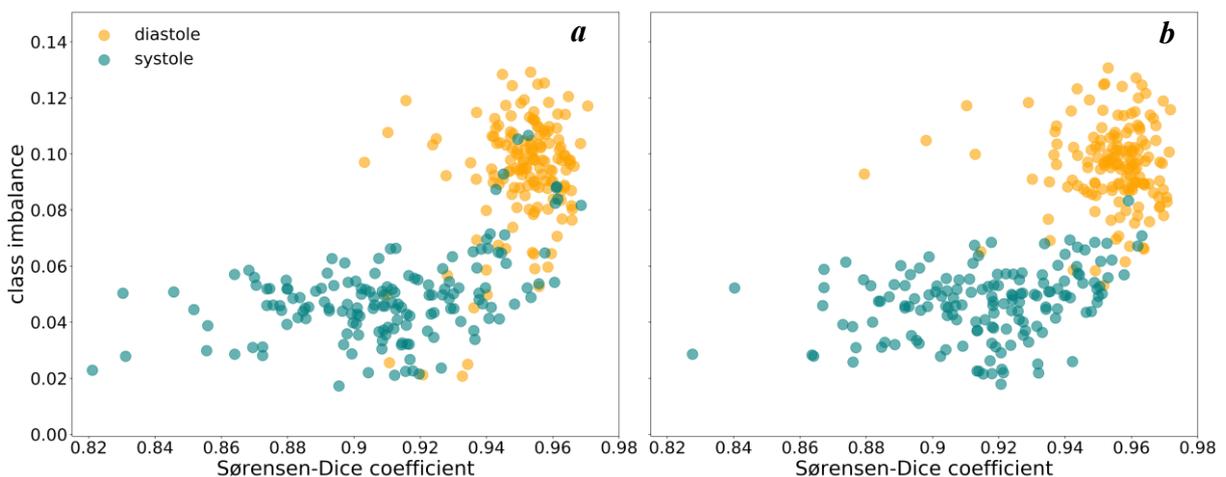

**Figure 8**. Class imbalance (foreground/background) vs Sørensen-Dice coefficient for systole and diastole phases for pooled Study 5 and Study 6 images with trained V-Net and Attention U-Net from 2MSA (*a*) and U-Net from 1MSA (*b*)

The hyperparameter tuning showed moderate performance improvement. Nonetheless, a few interesting facts were observed. Renormalisation deterred overfitting more effectively than direct regularisation techniques and this is in line with current knowledge within the imaging community. This claim is also corroborated by the substantial drop in DSC performance experienced whenever renormalisation was not implemented. Indeed, the application of dropout at random could deprive the model of its most critical training parameters and cause a performance erosion. Finally, the alleged superiority of activation functions with self-normalising properties (i.e. GELU and SELU)[37,38] was not observed in our context.

Data augmentation has the potential to improve performance,[7] but this was not the case for 1MSA, which could be due to the lack of capacity to incorporate instance variance. Further ablation studies could help better understanding this phenomenon.

In principle, it was expected that the 2MSA would lead to significantly better segmentations than 1MSA. However, the benchmarking of both 1MSA and 2MSA revealed that they were virtually equivalent. This is indeed supported by the LMM fit, the equivalence test and MD analyses. Both 1MSA and 2MSA demonstrated to model appropriately EFs in the range between 50 % and 70 %, whilst higher variability was observed outside this region. This likely stems from the natural limitation that the number of available training examples is lower.

The segmentation capability of these models was acceptable when compared to humans, as the bias and limits of agreement in the Bland-Altman plots were satisfactory (detecting an EF change of 5 % translates to clinically meaningful results on cardiac performance). Additionally, further agreement analysis revealed that these models held a promise in remediating inter- and intraoperator variability, as well as dramatically reducing the time of analysis. In line with previous efforts,[39] both 1MSA and 2MSA segmentation approaches had a lower bias than between operators and the limits of agreement were comparable. Certainly, these models while substantially quicker (~12 s with automated segmentation vs. ~20 mins with manual segmentation based on *in-house* experience). A shortcoming of this study is the limited number of operators available. Taking note of this, the addition of more operators for the assessment of the intraoperator agreement should be explored in the future.

Comparable levels of agreement between 1MSA and 2MSA prompted further investigation into the benefits of the former approach. Its applicability for automated phase selection leveraging the temporal dimension of each image stack along the cardiac cycle was explored. This approach expands on previous research carried out with 2D human images.[40] Amongst the phase selection metrics, the slice area afforded the lowest bias, potentially due to its sensitivity to small changes. The presence of bias in the selection metrics can be attributed to potential computational artifacts, biases in the data or in the annotation process. The surface area yielded the narrowest limits of agreement, which translates into better precision, as it has lower susceptibility to the noise potentially present in some slices. Thus far, no indication of a selection metric providing both an optimal precision and accuracy has been found. The addition of more rats into the study could help alleviate the bias which the surface area metric suffers from, whilst improving a high variability scenario would be more challenging. Therefore, using the surface area provides more reliable EF estimates than the slice area.

Next, the fitting method which was best suited to model the cardiac cycle was evaluated. Visual inspection of the GP and polynomial fits revealed that the GP ability to learn the distribution of datapoints, furnishes it with the desired flexibility needed to model the cardiac cycle. Additionally, GPs offer the possibility to develop more advanced approaches such as sparse and variational Gaussian Process[41] which can improve further the representation of the cardiac cycle.

The 1MSA provides a seamless phase selection which, coupled with estimation of the EF, allows for a full-automated process. This reduces the segmentation time dramatically and has potential to substantially accelerate pharmacological studies. Whilst 1MSA is beneficial in some contexts, 2MSA can be used in a context where sporadic human intervention is a possibility. It is a simpler approach which can prevent potential errors that might arise from the phase selection process. Thus, the suitability of using either 1MSA or 2MSA would be very much dependant on the specific requirements from the end-user.

Future work will focus around generalising these frameworks for new clinical studies and subsequently improving the predictive capabilities of the existing models by exploring continual learning[42,43] and transfer learning.[44,45]

The success of our methodology with a relatively low number of training instances is paves the way for its application to other organs and animals of preclinical interest. In this respect, preliminary work indicates that transfer learning could be applied successfully in the segmentation of mice hearts and we aim to explore this area further. Our work deals with a fundamental issues of image segmentation in preclinical settings and the success of our models trained on a relatively small preclinical dataset highlights the potential of this technique to be applied to automated cardiac segmentation both in preclinical settings.

**CONCLUSIONS**

We present a successful application of CNNs for the segmentation of rat LV from CMR datasets. Estimates of derived clinical parameters, such as systolic left-ventricular volume, diastolic left-ventricular volume, and EF, demonstrate close to human performance in both 2MSA and 1MSA settings. Combined with a novel phase selection strategy based on GPs modelled on the changes of the mid-slice are along the segmented cardiac cycle, our work presents an encouraging first step towards a fully automated rat LV segmentation pipeline. In the future, due to their potential to significantly speed up reporting of results to analysis teams, we hope that such tools will be useful when assessing pharmacological interventions designed to improve cardiac function. As such, we believe that the deep learning based automated segmentation pipeline presented here would inspire research into similar assistive technologies in the preclinical image analysis space in the years to come.

**SUPPLEMENTARY INFORMATION**

Segmentation Approach Comparison; Implementation Details; Model Selection Statistics; Segmentation Quality Analysis by Slice; Noise Robustness Analysis, Hyperparameter Tuning; Interoperator Volume Agreement; Bland-Altman Plots for Automation Feasibility Studies.

**LIST OF ABBREVIATIONS**

**1MSA:** One-Model Segmentation Approach

**2MSA:** Two-Model Segmentation Approach

**CI**: Confidence Interval

**CMR:** Cardiac Magnetic Resonance imaging

**CNN:** Convolutional Neural Networks

**DSC:** Sørensen-Dice Score

**ED:** End-Diastole

**EF:** Ejection Fraction

**ES:** End-Systole

**GP:** Gaussian Process

**ICC:** Intra-class Correlation

**LV:** Left Ventricle

**LOESS:** Locally estimated scatterplot smoothing

**LMM:** Linear Mixed Model

**MD**: Mean absolute Difference (univariate)

**RBF**: Radial-Basis Function

**ROI:** Region Of Interest

**SNR:** Signal to Noise Ratio

# DECLARATIONS

The authors would like to highlight the fact that this paper has not yet undergone peer review and therefore the findings presented are provisional and that the conclusions may change.

*Ethics approval and consent to participate*

All experiments were performed in compliance with EU Directive 2010/63/EU. The protocols were approved by the local ethics committee (Göteborgs djurförsöksetisk nämnd).

*Availability of data and materials*

The datasets generated and/or analysed during the current study are not publicly available due to company restrictions but are available from the corresponding author on reasonable request.

Code will be made available on request from the corresponding author.

*Competing interests*

DF, PKo, LH, PKa, AP and HV are employees of AstraZeneca and own stock options.

*Funding*

All authors were funded by Biopharmaceutical R&D, AstraZeneca Pepparedsleden 1, SE 43183 Mölndal, Sweden.

*Authors' information*

**Clinical Pharmacology and Safety Sciences, Biopharmaceutical R&D, AstraZeneca, Pepparedsleden 1, SE 431 83 Mölndal, Sweden**

D. Fernández-Llaneza (daniel.fernandez1@astrazeneca.com);

A. Gondová (ada.gondova@gmail.com);

A. Patra (arijit.patra@astrazeneca.com);

H. Vince (harris.vince@astrazeneca.com);

M. Zurek (magdalena.zurek-knab@novartis.com);

P. Kagelid (patrik.kagelid@astrazeneca.com);

L. Hultin(leif.hultin@astrazeneca.com).

**Data Sciences & Quantitative Biology, Discovery Sciences, Biopharmaceuticals R&D, AstraZeneca, Pepparedsleden 1, SE 431 83 Mölndal, Sweden**

P. Konings (peter.konings@astrazeneca.com).

*Authors' contributions*

DF, AG, MZ and LH carried out the conception and study design. MZ and LH performed the data collection and protocol design. DF and AG developed the algorithms, the analysis software and were responsible for the data analysis. DF and AG drafted the manuscript. All the authors reviewed and approved the manuscript.

*Corresponding author*

Correspondence to daniel.fernandez1@astrazeneca.com

*Acknowledgements*

The authors would like to thank Johan Karlsson, Edmund Watson, Tobias Noeske, Juan Pedro Vigueras-Guillén, Margareta Behrendt and Abdel Bidar for the fruitful discussions and help throughout the project.

**REFERENCES**

1.  Vandsburger MH, Epstein FH. Emerging MRI methods in translational cardiovascular research. *J Cardiovasc Transl Res*. 2011;4(4):477-492. doi:10.1007/s12265-011-9275-1

2.  Riehle C, Bauersachs J. Small animal models of heart failure. *Cardiovasc Res*. 2019;115(13):1838-1849. doi:10.1093/cvr/cvz161

3.  Caudron J, Fares J, Bauer F, Dacher J-N. Evaluation of left ventricular diastolic function with cardiac MR imaging. *Radiogr a Rev Publ Radiol Soc North Am Inc*. 2011;31(1):239-259. doi:10.1148/rg.311105049

4.  Miller CA, Jordan P, Borg A, et al. Quantification of left ventricular indices from SSFP cine imaging: impact of real-world variability in analysis methodology and utility of geometric modeling. *J Magn Reson Imaging*. 2013;37(5):1213-1222. doi:10.1002/jmri.23892

5.  Petitjean C, Dacher J-N. A review of segmentation methods in short axis cardiac MR images. *Med Image Anal*. 2011;15(2):169-184. doi:https://doi.org/10.1016/j.media.2010.12.004

6.  Xu H. Fully Automated Segmentation of the Left Ventricle in Small Animal Cardiac MRI. In: ; 2018.

7.  Hammouda K, Khalifa F, Abdeltawab H, et al. A New Framework for Performing Cardiac Strain Analysis from Cine MRI Imaging in Mice. *Sci Rep*. 2020;10:7725. doi:https://doi.org/10.1038/s41598-020-64206-x

8.  Zufiria B, Stephens M, Sánchez MJ, Ruiz-Cabello J, López-Linares K, Macía I. Fully Automatic Cardiac Segmentation And Quantification For Pulmonary Hypertension Analysis Using Mice Cine Mr Images. In: *2021 IEEE 18th International Symposium on Biomedical Imaging (ISBI)*. ; 2021:1411-1415. doi:10.1109/ISBI48211.2021.9433855

9.  Ronneberger O, Fischer P, Brox T. U-Net: Convolutional Networks for Biomedical Image Segmentation. *Lect Notes Comput Sci (including Subser Lect Notes Artif Intell Lect Notes Bioinformatics)*. 2015;9351:234-241. doi:10.1007/978-3-319-24574-4_28

10. Bernard O, Lalande A, Zotti C, et al. Deep Learning Techniques for Automatic MRI Cardiac Multi-Structures Segmentation and Diagnosis: Is the Problem Solved? *IEEE Trans Med Imaging*. 2018;37(11):2514-2525. doi:10.1109/TMI.2018.2837502

11. Çiçek Ö, Abdulkadir A, Lienkamp SS, Brox T, Ronneberger O. 3D U-net: Learning dense volumetric segmentation from sparse annotation. *Lect Notes Comput Sci (including Subser Lect Notes Artif Intell Lect Notes Bioinformatics)*. 2016;9901 LNCS:424-432. doi:10.1007/978-3-319-46723-8_49

12. Oktay O, Schlemper J, Folgoc L Le, et al. Attention U-Net: Learning Where to Look for the Pancreas. 2018;(Midl). http://arxiv.org/abs/1804.03999.

13. Zhou Z, Siddiquee MMR, Tajbakhsh N, Liang J. UNet++: Redesigning Skip Connections to Exploit Multiscale Features in Image Segmentation. *IEEE Trans Med Imaging*. 2020;39(6):1856-1867. doi:10.1109/TMI.2019.2959609

14. Zhou Z, Rahman Siddiquee MM, Tajbakhsh N, Liang J. Unet++: A nested u-net architecture for medical image segmentation. *Lect Notes Comput Sci (including Subser Lect Notes Artif Intell Lect Notes Bioinformatics)*. 2018;11045 LNCS:3-11. doi:10.1007/978-3-030-00889-5_1

15. Milletari F, Navab N, Ahmadi SA. V-Net: Fully convolutional neural networks for volumetric medical image segmentation. *Proc - 2016 4th Int Conf 3D Vision, 3DV 2016*. 2016:565-571. doi:10.1109/3DV.2016.79

16. Ruijsink B, Puyol-Antón E, Oksuz I, et al. Fully Automated, Quality-Controlled Cardiac Analysis From CMR: Validation and Large-Scale Application to Characterize Cardiac Function.

*JACC Cardiovasc Imaging*. 2020;13(3):684-695. doi:10.1016/j.jcmg.2019.05.030

17. Chen C, Qin C, Qiu H, et al. Deep learning for cardiac image segmentation: A review. *arXiv*. 2019;7(March). doi:10.3389/fcvm.2020.00025

18. Sonka M, Hlavac V, Boyle R. Mathematical morphology. In: *Image Processing, Analysis and Machine Vision*. Boston, MA: Springer US; 1993:422-442. doi:10.1007/978-1-4899-3216-7_10

19. Bergstra J, Bengio Y. Random search for hyper-parameter optimization. *J Mach Learn Res*. 2012;13:281-305.

20. Ioffe S. Batch Renormalization: Towards reducing minibatch dependence in batch-normalized models. *Adv Neural Inf Process Syst*. 2017;2017-Decem:1946-1954.

21. Dice LR. Measures of the Amount of Ecologic Association Between Species. *Ecology*. 1945;26(3):297-302. doi:https://doi.org/10.2307/1932409

22. Verbeke G, Molenberghs G. *Linear Mixed Models for Longitudinal Data*. Springer-Verlag New York; 2000. doi:10.1007/978-1-4419-0300-6

23. Hauck WW, Anderson S. A new statistical procedure for testing equivalence in two-group comparative bioavailability trials. *J Pharmacokinet Biopharm*. 1984;12(1):83-91. doi:10.1007/BF01063612

24. Gudbjartsson H, Patz S. The Rician Distribution of Noisy MRI Data. *Magn Reson Med*. 1996;36(2):332.

25. Edelstein WA, Bottomley PA, Pfeifer LM. A signal-to-noise calibration procedure for NMR imaging systems. *Med Phys*. 1984;11(2):180-185. doi:https://doi.org/10.1118/1.595484

26. Bland JM, Altman DG. Comparing methods of measurement: why plotting difference against standard method is misleading. *Lancet (London, England)*. 1995;346(8982):1085-1087. doi:10.1016/s0140-6736(95)91748-9

27. Rasmussen CE, Williams CK. *Gaussian Process for Machine Learning*. The MIT Press; 2006.

28. Lewiner T, Lopes H, Vieira AW, Tavares G. Efficient Implementation of Marching Cubes' Cases with Topological Guarantees. *J Graph Tools*. 2003;8(2):1-15. doi:10.1080/10867651.2003.10487582

29. Chollet F. Keras. 2019. https://keras.io/.

30. Abadi M, Agarwal A, Barham P, et al. TensorFlow: Large-Scale Machine Learning on Heterogeneous Distributed Systems. 2016. http://arxiv.org/abs/1603.04467.

31. Russell A, Lenth V, Buerkner P, et al. Package ' emmeans ' R topics documented : 2021;34(1):216-221. doi:10.1080/00031305.1980.10483031

32. R Core Team. R: A Language and Environment for Statistical Computing. 2020. https://www.r-project.org/.

33. Lenth R V., Buerkner P, Herve M, Love J, Riebl H, Singmann H. emeans: Estimates Marginal Means, aka Least-Squares Means. 2020. https://github.com/rvlenth/emmeans.

34. Bai W, Sinclair M, Tarroni G, et al. Automated cardiovascular magnetic resonance image analysis with fully convolutional networks. *arXiv*. 2017:1-12. doi:10.1186/s12968-018-0471-x

35. Sander J, de Vos BD, Išgum I. Automatic segmentation with detection of local segmentation failures in cardiac MRI. *Sci Rep*. 2020;10(1):1-19. doi:10.1038/s41598-020-77733-4

36. Poudel RPK, Lamata P, Montana G. Recurrent Fully Convolutional Neural Networks for Multi-slice MRI Cardiac Segmentation. In: Zuluaga MA, Bhatia K, Kainz B, Moghari MH, Pace DF, eds. *Reconstruction, Segmentation, and Analysis of Medical Images*. Cham: Springer International Publishing; 2017:83-94.

37. Hendrycks D, Gimpel K. Gaussian Error Linear Units (GELUs). 2016:1-9. http://arxiv.org/abs/1606.08415.

38. Unterthiner T, Hochreiter S. Self-Normalizing Neural Networks. 2017;(Nips).

39. Riegler J, Cheung KK, Man YF, Cleary JO, Price AN, Lythgoe MF. Comparison of Segmentation Methods for MRI Measurement of Cardiac Function in Rats. *J Magn Reson Imaging*. 2010;32(4):869-877. doi:10.1002/jmri.22305

40. Kong B, Zhan Y, Shin M, Denny T, Zhang S. Recognizing End-Diastole and End-Systole Frames via Deep Temporal Regression Network. In: Ourselin S, Joskowicz L, Sabuncu MR, Unal G, Wells W, eds. *Medical Image Computing and Computer-Assisted Intervention - MICCAI 2016*. Cham: Springer International Publishing; 2016:264-272.

41. de G. Matthews AG, Hensman J, Turner RE, Ghahramani Z. On Sparse variational methods and the Kullback-Leibler divergence between stochastic processes. 2015.

42. Thrun S. Is Learning The n-th Thing Any Easier Than Learning The First? *Adv Neural Inf Process Syst*. 1996:7. http://citeseer.ist.psu.edu/viewdoc/summary?doi=10.1.1.44.2898.

43. Thrun S. Explanation-Based Neural Network Learning. In: *Explanation-Based Neural Network Learning: A Lifelong Learning Approach*. Boston, MA: Springer US; 1996:19-48. doi:10.1007/978-1-4613-1381-6_2

44. Weiss K, Khoshgoftaar TM, Wang DD. *A Survey of Transfer Learning*. Vol 3. Springer International Publishing; 2016. doi:10.1186/s40537-016-0043-6

45. Pan SJ, Yang Q. A Survey on Transfer Learning. *IEEE Trans Knowl Data Eng*. 2010;22(10):1345-1359. doi:10.1109/TKDE.2009.191

# Supplementary Information

# Towards Fully Automated Segmentation of Rat Cardiac MRI by Leveraging Deep Learning Frameworks

Daniel Fernández-Llaneza,[*,†,§] Andrea Gondová,[†, §] Harris Vince,[†] Arijit Patra,[†] Magdalena Zurek,[†] Peter Konings,[‡] Patrik Kagelid,[†] Leif Hultin[†]

[†] Clinical Pharmacology and Safety Sciences, Biopharmaceuticals R&D, AstraZeneca, Pepparedsleden 1, SE 431 83 Mölndal, Sweden

[‡] Data Sciences & Quantitative Biology, Discovery Sciences, Biopharmaceuticals R&D, AstraZeneca, Pepparedsleden 1, SE 431 83 Mölndal, Sweden

[§] DF and AG contributed equally to this work

___________________

[*] e-mail: daniel.fernandez1@astrazeneca.com (D. Fernández-Llaneza)

# Contents

# 1. Segmentation Approach Comparison
## 1.1. Linear Mixed Model
### 1.1.1. Implementation

These variables are expressed as indicator functions as such:

$$\mathbb{I}_{reader}(x_i) := \begin{cases} 1 & \text{if } x_i \in \text{Op2} \\ 0 & \text{if } x_i \in \text{Op1} \end{cases}$$

$$\mathbb{I}_{approach}(x_i) := \begin{cases} 1 & \text{if } x_i \in \text{2MSA} \\ 0 & \text{if } x_i \in \text{1MSA} \end{cases}$$

$$\mathbb{I}_{treatment}(x_i) := \begin{cases} 1 & \text{if } x_i \in \text{naïve} \\ 0 & \text{if } x_i \in \text{myocardial infarction} \end{cases} \qquad \text{equation S1}$$

### 1.1.2. Linear Mixed Model Diagnostics

The QQ plot corroborates that the mixed linear model residuals are normally distributed, although there is some deviation from the expected normal line towards the tails which could indicate kurtosis. Nonetheless, the residual distribution is good enough, so the assumption of normality is tenable.

## 1.2. LOESS fit

A LOESS fit was used as an exploratory analysis tool to identify trends in manual and automated segmentations (see **Fig. S1**). Naïve rats have high ejection fractions (> 55 %), while there is a higher degree of variability in the case of myocardial infarction rats whose EFs span a range from 30 % up to 70 %, approximately. The modelled EFs deviate from linearity in the extremes. For Op1, the diagonal switches sides at around the 50 % EF mark. This trend is also identified for Op2, albeit it is less pronounced. No major differences are observed when comparing the fits between the two segmentation approaches.

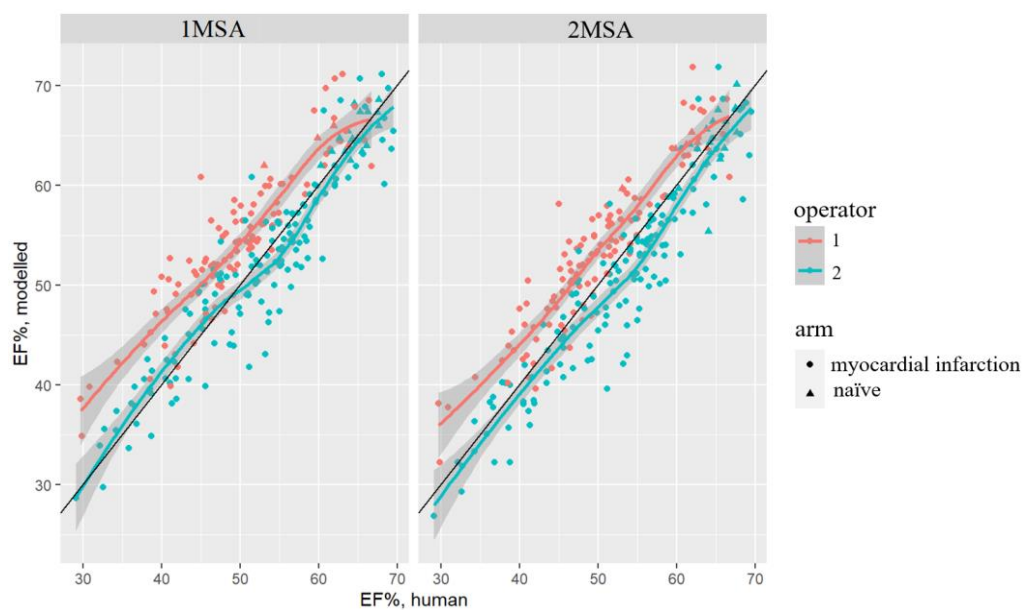

**Figure S1.** LOESS fit for EF determined using automated segmentation vs. manual segmentation by segmentation approach (*left*: one-model segmentation approach, *right*: two-model segmentation approach).
Both graphs plot the operators (— Operator 1, — Operator 2) involved in the segmentation and the treatment arm (● myocardial infarction, ▲ naïve)

# 2. Implementation Details
## 2.1. Gaussian Process

The crux of GP lies in defining a kernel which establishes the prior covariance matrix between two function values *x* and *x'* such:

$$k : \mathbb{R}^n \times \mathbb{R}^n \rightarrow \mathbb{R}, \quad \Sigma = \text{Cov}[f(\mathbf{X}), f(\mathbf{X'})] = k(\mathbf{x},\mathbf{x'}) \qquad \text{equation S2}$$

The composite kernel consisting of a constant kernel multiplied by the radial-basis function (RBF) kernel is given by:

$$k(\mathbf{x},\mathbf{x'}) = \text{constant value} \quad \forall \; \mathbf{x}, \mathbf{x'} \qquad \text{equation S3}$$

$$k(\mathbf{x},\mathbf{x'}) = \exp\left(\frac{d(\mathbf{x},\mathbf{x'})^2}{2l^2}\right) \qquad \text{equation S4}$$

where the constant value for the constant kernel is set to 0.1 with constant value bounds between 0.1 and 10, $d(\cdot, \cdot)$ is the Euclidean distance, $l$ is a length-scale parameter which is set to 0.5 with bounds 0.1 to 10.

# 3. Model Selection Statistics

**Table S1. DSC Statistic for Final Models in 2MSA**[a]

| model | phase model | Study 5 | Study 6 |
|---|---|---|---|
| Attention U-Net | systole | 0.90 ± 0.024 | 0.93 ± 0.019 |
|  | **diastole** | **0.95 ± 0.014** | **0.96 ± 0.011** |
| U-Net | systole | 0.90 ± 0.23 | 0.94 ± 0.017 |
| U-Net++ | diastole | 0.94 ± 0.013 | 0.90 ± 0.016 |
| V-Net | **systole** | **0.91 ± 0.022** | **0.94 ± 0.015** |
|  | diastole | 0.95 ± 0.013 | 0.95 ± 0.00087 |
| voting ensemble | systole | 0.91 ± 0.023 | 0.94 ± 0.015 |
|  | diastole | 0.95 ± 0.0095 | 0.86 ± 0.016 |
| averaging ensemble | systole | 0.89 ± 0.027 | 0.93 ± 0.023 |
|  | diastole | 0.95 ± 0.012 | 0.90 ± 0.015 |

[a] DSC is reported as mean ± standard deviation

**Table S2. DSC Statistic for Final Models in 1MSA**[a]

| model | phase | Study 5 | Study 6 |
|---|---|---|---|
| Attention U-Net | systole | 0.90 ± 0.026 | 0.94 ± 0.019 |
|  | diastole | 0.95 ± 0.013 | 0.95 ± 0.014 |
| **U-Net** | **systole** | **0.90 ± 0.024** | **0.94 ± 0.018** |
|  | **diastole** | **0.95 ± 0.011** | **0.95 ± 0.014** |
| V-Net | systole | 0.90 ± 0.024 | 0.94 ± 0.025 |
|  | diastole | 0.95 ± 0.013 | 0.95 ± 0.021 |
| voting ensemble | systole | 0.90 ± 0.024 | 0.94 ± 0.019 |
|  | diastole | 0.96 ± 0.013 | 0.96 ± 0.0079 |
| averaging ensemble | systole | 0.88 ± 0.030 | 0.94 ± 0.025 |
|  | diastole | 0.95 ± 0.015 | 0.96 ± 0.016 |

[a] DSC is reported as mean ± standard deviation

The Hausdorff Distance (HD) from A to B is defined as follows:

$$\tilde{\delta}_H(A,B) = \max_{a \in A} \min_{b \in B} \|a-b\| \qquad \text{Equation S4}$$

and subsequently, the bidirectional Hausdorff distance is calculated as specified in *Equation 2*.

$$HD(A,B) = \max(\tilde{\delta}_H(A,B), \tilde{\delta}_H(B,A)) \qquad \text{Equation S5}$$

Intraclass Correlation Coefficient (ICC) was determined to evaluate agreement between the actual volumes (derived from the true masks) and the predicted volumes. A value of 1 indicates perfect reliability, whereas a value of 0 indicates no agreement.

**Table S3. HD Statistic for Final Models in 2MSA**[a]

| model | phase model | Study 5 | Study 6 |
|---|---|---|---|
| Attention U-Net | systole | 2.0 ± 0.45 | 1.6 ± 0.31 |
|  | diastole | 1.63 ± 0.27 | 1.7 ± 1.1 |
| U-Net | systole | 2.0 ± 0.41 | 1.6 ± 0.27 |
| U-Net++ | diastole | 1.83 ± 0.50 | 2.0 ± 0.97 |
| V-Net | systole | 2.1 ± 0.44 | 1.59 ± 0.13 |
|  | diastole | 1.61 ± 0.28 | 1.8 ± 1.2 |
| voting ensemble | systole | 1.9 ± 0.40 | 1.52 ± 0.18 |
|  | diastole | 1.6 ± 0.17 | 2.12 ± 0.95 |
| averaging ensemble | systole | 2.1 ± 0.48 | 1.64 ± 0.23 |
|  | diastole | 1.79 ± 0.48 | 1.9 ± 0.98 |

[a] HD is reported as mean ± standard deviation

**Table S4. ICC Statistic for Final Models in 2MSA**[a]

| model | phase model | Study 5 | Study 6 |
|---|---|---|---|
| Attention U-Net | systole | 0.79 (0.71 to 0.85) | 0.98 (0.96 to 0.99) |
|  | diastole | 0.90 (0.86 to 0.93) | 0.96 (0.92 to 0.98) |
| U-Net | systole | 0.79 (0.71 to 0.85) | 0.99 (0.97 to 0.99) |
| U-Net++ | diastole | 0.92 (0.88 to 0.94) | 0.59 (0.34 to 0.76) |
| V-Net | systole | 0.84 (0.78 to 0.89) | 0.99 (0.98 to 0.99) |
|  | diastole | 0.88 (0.84 to 0.92) | 0.92 (0.86 to 0.96) |
| voting ensemble | systole | 0.81 (0.73 to 0.86) | 0.99 (0.98 to 0.99) |
|  | diastole | 0.93 (0.90 to 0.95) | 0.30 (0.000 to 0.57) |
| averaging ensemble | systole | 0.66 (0.54 to 0.75) | 0.96 (0.93 to 0.98) |
|  | diastole | 0.92 (0.88 to 0.94) | 0.62 (0.38 to 0.78) |

[a] ICC volume as mean and 95% confidence interval in brackets

**Table S5. HD Statistic for Final Models in 1MSA**[a]

| model | phase model | Study 5 | Study 6 |
|---|---|---|---|
| Attention U-Net | systole | 2.0 ± 0.38 | 1.7 ± 0.52 |
|  | diastole | 1.6 ± 0.32 | 1.7 ± 1.2 |
| U-Net | systole | 2.0 ± 0.48 | 1.6 ± 0.20 |
|  | diastole | 1.6 ± 0.26 | 1.8 ± 1.2 |
| V-Net | systole | 2.1 ± 0.47 | 1.6 ± 0.20 |
|  | diastole | 1.6 ± 0.31 | 1.8 ± 1.2 |
| voting ensemble | systole | 2.0 ± 0.46 | 1.6 ± 0.17 |
|  | diastole | 1.6 ± 0.28 | 1.7 ± 1.2 |
| averaging ensemble | systole | 2.2 ± 0.49 | 1.7 ± 0.23 |
|  | diastole | 1.7 ± 0.33 | 1.7 ± 1.2 |

[a] HD is reported as mean ± standard deviation

**Table S6. ICC Statistic for Final Models in 1MSA**[a]

| model | phase model | Study 5 | Study 6 |
|---|---|---|---|
| Attention U-Net | systole | 0.79 (0.70 to 0.84) | 0.98 (0.97 to 0.99) |
|  | diastole | 0.91 (0.87 to 0.94) | 0.95 (0.90 to 0.97) |
| U-Net | systole | 0.80 (0.72 to 0.85) | 0.98 (0.95 to 0.99) |
|  | diastole | 0.91 (0.88 to 0.94) | 0.94 (0.88 to 0.97) |
| V-Net | systole | 0.79 (0.71 to 0.85) | 0.97 (0.95 to 0.99) |
|  | diastole | 0.94 (0.91 to 0.96) | 0.95 (0.91 to 0.97) |
| voting ensemble | systole | 0.79 (0.71 to 0.85) | 0.98 (0.96 to 0.99) |
|  | diastole | 0.92 (0.88 to 0.94) | 0.95 (0.90 to 0.97) |
| averaging ensemble | systole | 0.62 (0.49 to 0.72) | 0.95 (0.91 to 0.97) |
|  | diastole | 0.81 (0.74 to 0.86) | 0.88 (0.78 to 0.93) |

[a] ICC volume as mean and 95% confidence interval in brackets

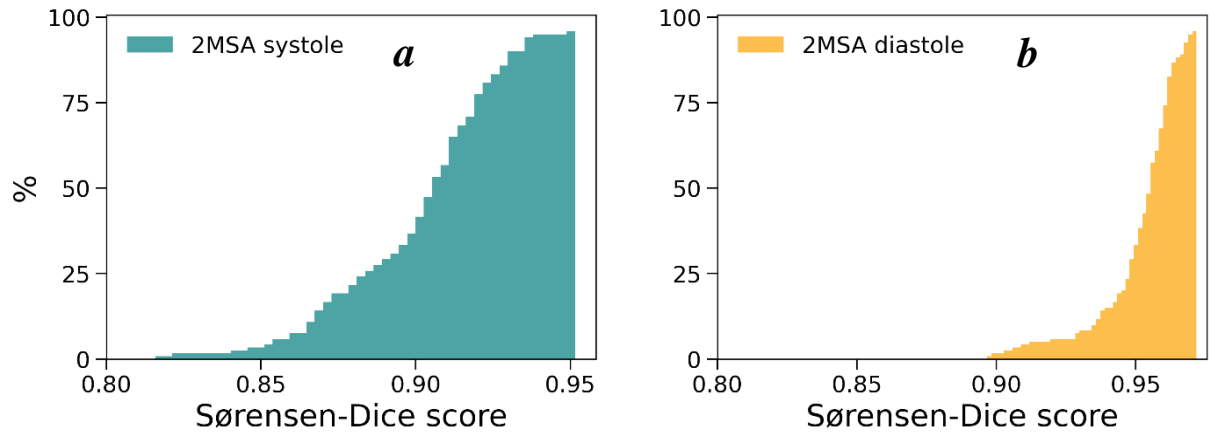

**Figure S2.** Sørensen-Dice score cumulative distribution for 2MSA.
*a*: V-Net
*b*: Attention U-Net

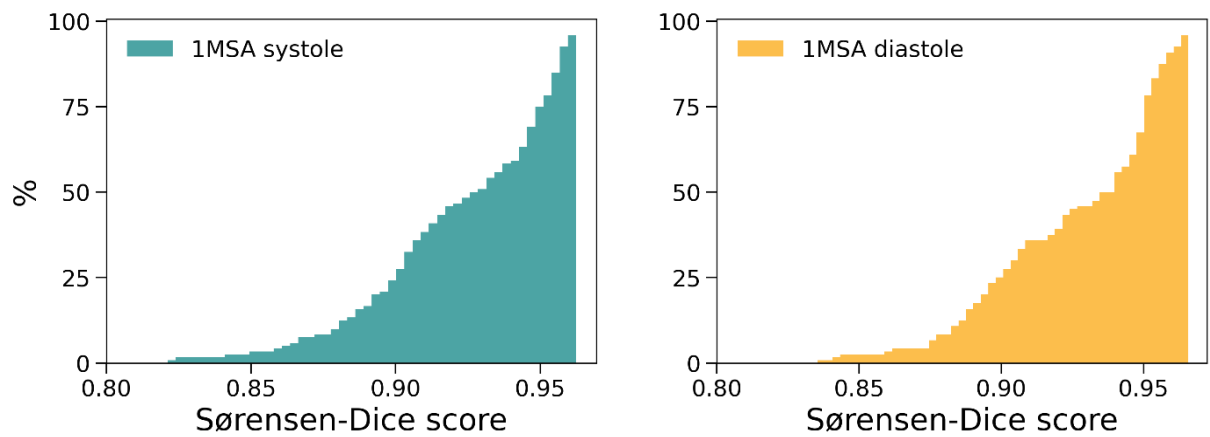

**Figure S3.** Sørensen-Dice score cumulative distribution for 1MSA U-Net model
Supplementary Information, S5/S21

## 4. Segmentation Quality Analysis by Slice

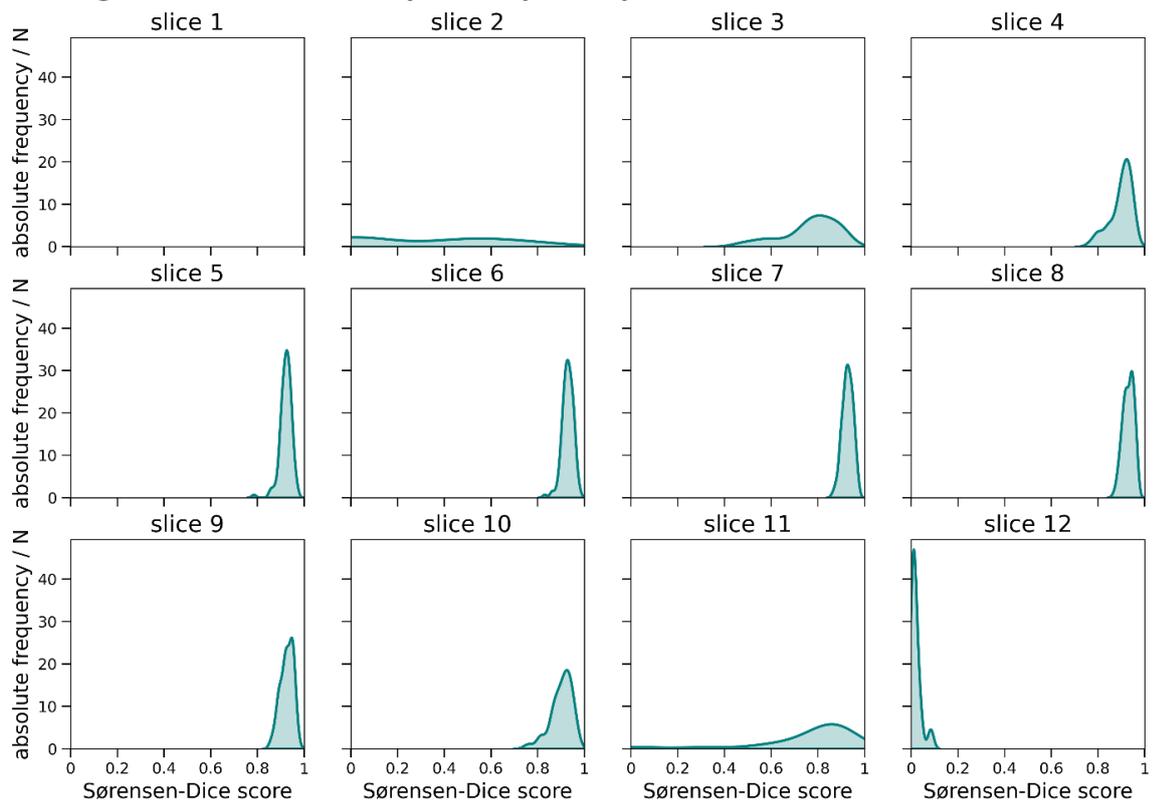

**Figure S4.** Sørensen-Dice score distribution per slice for systolic phase segmentations, where the slice 1 is the basal slice and the last slice the apical slice.

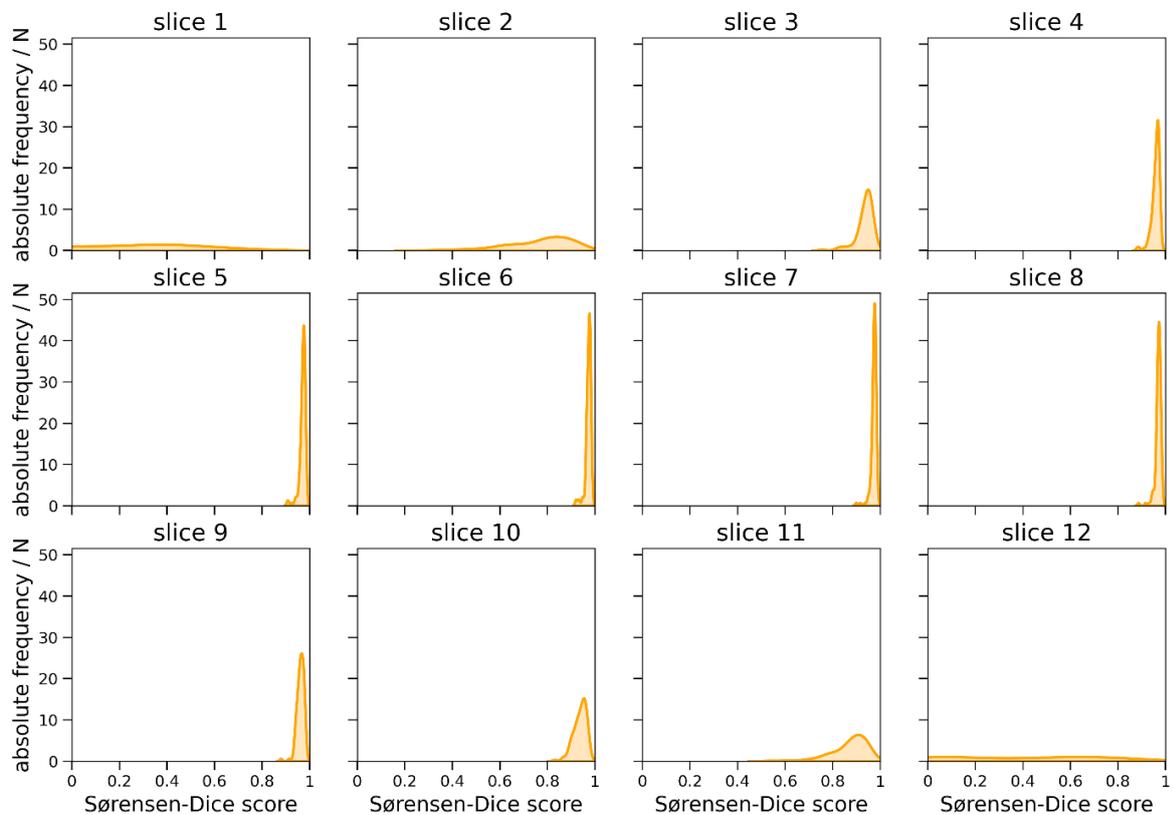

**Figure S5**. Sørensen-Dice score distribution per slice for diastolic phase segmentations, where the slice 1 is the basal slice and the last slice the apical slice.

# 5. Noise Robustness Analysis

**Table S7. Noise Robustness Analysis for Two-model Segmentation Approach**

| approach | architecture | phase | metric | model type | Gaussian noise | Rician noise | Rayleigh noise | mixed noise |
|---|---|---|---|---|---|---|---|---|
| two-model segmentation approach | Attention U-Net | diastole | HD | non-augmented | 5.8 ± 3.9 | 6.0 ± 3.8 | 8.9 ± 3.4 | 12.0 ± 7.4 |
| | | | | augmented | 1.6 ± 0.27 | 1.6 ± 0.28 | 1.6 ± 0.27 | 1.7 ± 0.41 |
| | | | ICC | non-augmented | 0.75 (0.66 to 0.82) | 0.72 (0.62 to 0.80) | 0.27 (0.093 to 0.43) | 0.26 (0.086 to 0.42) |
| | | | | augmented | 0.90 (0.86 to 0.94) | 0.89 (0.84 to 0.92) | 0.88 (0.84 to 0.92) | 0.83 (0.77 to 0.88) |
| | V-Net | systole | HD | non-augmented | 2.1 ± 0.44 | 2.1 ± 0.46 | 2.1 ± 0.45 | 2.2 ± 0.47 |
| | | | | augmented | 1.9 ± 0.35 | 1.9 ± 0.35 | 2.0 ± 0.34 | 2.0 ± 0.40 |
| | | | ICC | non-augmented | 0.70 (0.60 to 0.78) | 0.81 (0.74 to 0.87) | 0.80 (0.72 to 0.86) | 0.74 (0.64 to 0.81) |
| | | | | augmented | 0.84 (0.78 to 0.89) | 0.83 (0.77 to 0.88) | 0.82 (0.76 to 0.88) | 0.80 (0.72 to 0.85) |

**Table S8. Noise Robustness Analysis for One-model Segmentation Approach**

| approach | architecture | phase | metric | model type | Gaussian noise | Rician noise | Rayleigh noise | mixed noise |
|---|---|---|---|---|---|---|---|---|
| one-model segmentation approach | U-Net | diastole | HD | non-augmented | 1.7 ± 0.30 | 1.7 ± 0.31 | 1.6 ± 0.27 | 1.8 ± 0.50 |
| | | | | augmented | 1.8 ± 0.42 | 2.0 ± 0.50 | 1.6 ± 0.29 | 2.1 ± 0.54 |
| | | | ICC | non-augmented | 0.88 (0.84 to 0.92) | 0.88 (0.83 to 0.91) | 0.91 (0.87 to 0.93) | 0.88 (0.84 to 0.92) |
| | | | | augmented | 0.91 (0.87 to 0.94) | 0.90 (0.86 to 0.93) | 0.90 (0.85 to 0.0.927) | 0.87 (0.82 to 0.91) |
| | | systole | HD | non-augmented | 2.2 ± 0.49 | 2.2 ± 0.48 | 2.2 ± 0.48 | 2.8 ± 1.6 |
| | | | | augmented | 1.9 ± 0.44 | 1.6 ± 0.26 | 2.0 ± 0.51 | 1.6 ± 0.30 |
| | | | ICC | non-augmented | 0.79 (0.71 to 0.85) | 0.79 (0.71 to 0.85) | 0.76 (0.67 to 0.83) | 0.73 (0.64 to 0.81) |
| | | | | augmented | 0.77 (0.69 to 0.84) | 0.77 (0.68 to 0.83) | 0.76 (0.68 to 0.83) | 0.70 (0.59 to 0.78) |

# 6. Hyperparameter Tuning

**Table S9. Blocks, layers and downsampling dimensions hyperparameter tuning for systole model (two-model segmentation approach)** [a]

| architecture | blocks | layers | downsampling dimensions | DSC |
|---|---|---|---|---|
| Attention U-Net | 2 | 4 | 3 | 0.91449 |
| Attention U-Net | 2 | 2 | 2 | 0.90946 |
| Attention U-Net | 4 | 4 | 3 | 0.90003 |
| Attention U-Net | 3 | 4 | 2 | 0.9102 |
| Attention U-Net | 4 | 6 | 3 | 0.89254 |
| Attention U-Net | 2 | 4 | 3 | 0.91517 |
| Attention U-Net | 3 | 8 | 2 | 0.90992 |
| Attention U-Net | 2 | 8 | 3 | 0.91959 |
| **Attention U-Net** | **3** | **2** | **3** | **0.92306** |
| Attention U-Net | 3 | 2 | 2 | 0.91162 |
| U-Net | 3 | 2 | 2 | 0.91095 |
| U-Net | 5 | 8 | 3 | 0.92392 |
| U-Net | 4 | 4 | 2 | 0.91780 |
| U-Net | 4 | 6 | 3 | 0.69443 |
| U-Net | 3 | 4 | 3 | 0.91719 |
| U-Net | 2 | 2 | 3 | 0.219913 |
| U-Net | 2 | 4 | 3 | 0.50917 |
| **U-Net** | **5** | **2** | **2** | **0.91466** |
| U-Net | 2 | 8 | 3 | 0.91249 |
| U-Net | 4 | 8 | 2 | 0.90594 |
| V-Net | 2 | 2 | 3 | 0.91436 |
| V-Net | 2 | 6 | 2 | 0.92544 |
| V-Net | 3 | 2 | 2 | 0.92006 |
| V-Net | 4 | 4 | 3 | 0.9205 |
| V-Net | 3 | 6 | 3 | 0.91475 |
| **V-Net** | **4** | **6** | **2** | **0.92596** |
| V-Net | 2 | 8 | 3 | 0.86622 |
| V-Net | 4 | 8 | 2 | 0.91496 |
| V-Net | 3 | 8 | 2 | 0.90558 |
| V-Net | 4 | 2 | 2 | 0.92054 |

[a] The model with the highest DSC in each of the architectures tried is highlighted in bold

**Table S10. Hyperparameter tuning for systole model (two-model segmentation approach)** [a]

| architecture (blocks, layers, dimensions) | kernel size | number of filters | renormalisation | activation function | dropout | batch size | optimiser | learnrate | pooling | kernel initialiser | deconvolution | DSC |
|---|---|---|---|---|---|---|---|---|---|---|---|---|
| Attention U-Net (3,2,3) | (5, 5, 5) | 32 | True | ELU | 0.2 | 32 | Adam | $10^{-3}$ | average | random normal | False | 0.91305 |
| Attention U-Net (3,2,3) | (2, 2, 2) | 8 | True | SELU | 0.4 | 16 | Adam | $10^{-3}$ | average | random normal | False | 0.63149 |
| Attention U-Net (3,2,3) | (3, 3, 3) | 16 | True | SELU | 0 | 64 | Adam | $10^{-3}$ | max | Glorot uniform | True | 0.87994 |
| Attention U-Net (3,2,3) | (5, 5, 5) | 32 | False | GELU | 0.4 | 32 | Adam | $10^{-3}$ | average | Glorot uniform | True | 0.07667 |
| Attention U-Net (3,2,3) | (5, 5, 5) | 16 | True | SELU | 0.2 | 32 | Adam | $10^{-2}$ | max | Glorot uniform | True | 0.91216 |
| Attention U-Net (3,2,3) | (2, 2, 2) | 8 | True | GELU | 0.2 | 64 | Adam | $10^{-3}$ | average | Glorot normal | False | 0.6799 |
| Attention U-Net (3,2,3) | (3, 3, 3) | 16 | False | GELU | 0.6 | 16 | RMSProp | $10^{-2}$ | average | Glorot normal | False | 0.08205 |
| Attention U-Net (3,2,3) | (5, 5, 5) | 16 | False | SELU | 0.4 | 16 | RMSProp | $10^{-3}$ | average | Glorot normal | True | 0.40656 |
| Attention U-Net (3,2,3) | (2, 2, 2) | 16 | True | GELU | 0.2 | 32 | Adam | $10^{-2}$ | max | glorot normal | False | 0.90011 |
| Attention U-Net (3,2,3) | (2, 2, 2) | 32 | False | ELU | 0.6 | 32 | Adam | $10^{-2}$ | max | glorot normal | False | 0.59155 |
| Attention U-Net (3,2,3) | (2, 2, 2) | 32 | True | SELU | 0.4 | 16 | Adam | $10^{-3}$ | max | Glorot normal | True | 0.77934 |
| Attention U-Net (3,2,3) | (3, 3, 3) | 8 | False | GELU | 0.4 | 64 | RMSProp | $10^{-2}$ | max | Glorot normal | False | 0.08081 |

*(continuation)*

| architecture (blocks, layers, dimensions) | kernel size | number of filters | renormalisation | activation function | dropout | batch size | optimiser | learnrate | pooling | kernel initialiser | deconvolution | DSC |
|---|---|---|---|---|---|---|---|---|---|---|---|---|
| Attention U-Net (3,2,3) | (5, 5, 5) | 8 | True | swish | 0.0 | 8 | Adam | $10^{-2}$ | average | Glorot normal | False | 0.92118 |
| Attention U-Net (3,2,3) | (5, 5, 5) | 8 | True | GELU | 0.6 | 32 | Adam | $10^{-2}$ | max | random normal | False | 0.64864 |
| Attention U-Net (3,2,3) | (5, 5, 5) | 16 | False | GELU | 0.4 | 32 | Adam | $10^{-2}$ | average | Glorot normal | False | 0.10512 |
| Attention U-Net (3,2,3) | (3, 3, 3) | 32 | True | ELU | 0.2 | 16 | RMSProp | $10^{-2}$ | max | Glorot normal | False | 0.79341 |
| Attention U-Net (3,2,3) | (3, 3, 3) | 8 | True | ELU | 0.2 | 16 | Adam | $10^{-3}$ | max | Glorot uniform | True | 0.75997 |
| Attention U-Net (3,2,3) | (5, 5, 5) | 8 | False | GELU | 0.6 | 8 | Adam | $10^{-3}$ | max | random normal | False | 0.01196 |
| Attention U-Net (3,2,3) | (3, 3, 3) | 16 | True | swish | 0.6 | 64 | Adam | $10^{-2}$ | average | Glorot normal | False | 0.27207 |
| Attention U-Net (3,2,3) | (3, 3, 3) | 16 | True | ELU | 0 | 8 | Adam | $10^{-2}$ | max | Glorot normal | True | 0.92306 |
| Attention U-Net (3,2,3) | (3, 3, 3) | 16 | True | ELU | 0 | 64 | Adam | $10^{-2}$ | max | Glorot normal | True | 0.89907 |
| Attention U-Net (3,2,3) | (3, 3, 3) | 16 | True | ELU | 0.2 | 8 | Adam | $10^{-2}$ | max | Glorot normal | True | 0.8925 |
| Attention U-Net (3,2,3) | (3, 3, 3) | 16 | True | swish | 0 | 8 | Adam | $10^{-2}$ | max | Glorot normal | True | 0.33729 |
| Attention U-Net (3,2,3) | (3, 3, 3) | 16 | True | ELU | 0 | 8 | Adam | $10^{-2}$ | max | Glorot normal | False | 0.89186 |
| Attention U-Net (3,2,3) | (3, 3, 3) | 16 | True | GELU | 0 | 8 | Adam | $10^{-2}$ | max | Glorot normal | True | 0.89368 |
| U-Net (5,2,2) | (5, 5, 5) | 8 | True | swish | 0 | 8 | Adam | $10^{-2}$ | max | Glorot normal | False | 0.89443 |
| U-Net (5,2,2) | (5, 5, 5) | 8 | True | swish | 0 | 32 | Adam | $10^{-2}$ | average | Glorot normal | False | 0.89288 |
| U-Net (5,2,2) | (5, 5, 5) | 8 | True | swish | 0 | 8 | Adam | $10^{-2}$ | average | Glorot normal | True | 0.89546 |

*(continuation)*

| architecture (blocks, layers, dimensions) | kernel size | number of filters | renormalisation | activation function | dropout | batch size | optimiser | learnrate | pooling | kernel initialiser | deconvolution | DSC |
|---|---|---|---|---|---|---|---|---|---|---|---|---|
| U-Net (5,2,2) | (5, 5, 5) | 8 | True | GELU | 0 | 8 | Adam | $10^{-2}$ | average | Glorot normal | False | 0.89465 |
| U-Net (5,2,2) | (3, 3, 3) | 16 | True | ELU | 0 | 8 | Adam | $10^{-2}$ | max | Glorot normal | True | 0.92306 |
| U-Net (5,2,2) | (3, 3, 3) | 16 | False | SELU | 0 | 16 | RMSProp | $10^{-3}$ | max | Glorot normal | True | 0.28012 |
| U-Net (5,2,2) | (3, 3, 3) | 16 | True | SELU | 0.2 | 8 | Adam | $10^{-3}$ | max | Glorot uniform | True | 0.90423 |
| U-Net (5,2,2) | (2, 2, 2) | 8 | True | ELU | 0 | 16 | Adam | $10^{-2}$ | average | random normal | True | 0.90533 |
| U-Net (5,2,2) | (3, 3, 3) | 16 | False | swish | 0.2 | 8 | Adam | $10^{-2}$ | max | Glorot uniform | False | 0.93156 |
| U-Net (5,2,2) | (5, 5, 5) | 8 | True | GELU | 0.2 | 32 | Adam | $10^{-2}$ | average | random normal | False | 0.01196 |
| U-Net (5,2,2) | (3, 3, 3) | 16 | False | ELU | 0.4 | 32 | RMSProp | $10^{-2}$ | max | Glorot uniform | True | 0.48532 |
| U-Net (5,2,2) | (5, 5, 5) | 32 | True | swish | 0.6 | 8 | RMSProp | $10^{-2}$ | average | Glorot uniform | True | 0.08121 |
| U-Net (5,2,2) | (5, 5, 5) | 8 | True | GELU | 0.4 | 16 | Adam | $10^{-2}$ | average | Glorot uniform | False | 0.92159 |
| U-Net (5,2,2) | (3, 3, 3) | 16 | True | SELU | 0.2 | 64 | RMSProp | $10^{-3}$ | max | Glorot normal | True | 0.16392 |
| U-Net (5,2,2) | (3, 3, 3) | 16 | False | SELU | 0.6 | 64 | RMSProp | $10^{-3}$ | max | Glorot uniform | True | 0.29401 |
| U-Net (5,2,2) | (5, 5, 5) | 32 | True | GELU | 0.2 | 16 | RMSProp | $10^{-2}$ | max | Glorot uniform | True | 0.88553 |
| U-Net (5,2,2) | (2, 2, 2) | 8 | False | ELU | 0 | 32 | Adam | $10^{-3}$ | average | Glorot normal | False | 0.53106 |
| U-Net (5,2,2) | (5, 5, 5) | 16 | False | ELU | 0.2 | 8 | RMSProp | $10^{-3}$ | max | Glorot uniform | False | 0.6165 |
| U-Net (5,2,2) | (3, 3, 3) | 16 | False | GELU | 0.2 | 8 | Adam | $10^{-2}$ | max | Glorot uniform | False | 0.65958 |
| U-Net (5,2,2) | (3, 3, 3) | 16 | True | swish | 0.2 | 8 | Adam | $10^{-2}$ | max | Glorot uniform | False | 0.9153 |
| U-Net (5,2,2) | (3, 3, 3) | 16 | False | swish | 0 | 8 | Adam | $10^{-2}$ | max | Glorot uniform | False | 0.07589 |
| V-Net (4,6,2) | (2, 2, 2) | 16 | True | GELU | 0.6 | 32 | Adam | $10^{-3}$ | max | random normal | True | 0.4701 |
| V-Net (4,6,2) | (3, 3, 3) | 32 | True | GELU | 0.6 | 16 | RMSProp | $10^{-3}$ | average | Glorot normal | True | 0.07824 |
| V-Net (4,6,2) | (2, 2, 2) | 8 | False | GELU | 0.4 | 8 | RMSProp | $10^{-2}$ | average | random normal | True | 0.07858 |
| V-Net (4,6,2) | (3, 3, 3) | 8 | False | ELU | 0.4 | 32 | Adam | $10^{-3}$ | max | Glorot uniform | True | 0.17348 |
| V-Net (4,6,2) | (2, 2, 2) | 16 | True | GELU | 0.2 | 32 | Adam | $10^{-2}$ | max | Glorot normal | True | 0.90206 |
| V-Net (4,6,2) | (2, 2, 2) | 16 | True | SELU | 0.6 | 32 | RMSProp | $10^{-3}$ | average | random normal | True | 0.0779 |
| V-Net (4,6,2) | (2, 2, 2) | 32 | True | swish | 0.2 | 32 | Adam | $10^{-3}$ | max | Glorot normal | True | 0.86483 |
| V-Net (4,6,2) | (2, 2, 2) | 8 | True | ELU | 0.4 | 16 | Adam | $10^{-3}$ | average | Glorot normal | True | 0.36272 |

*(continuation)*

| architecture (blocks, layers, dimensions) | kernel size | number of filters | renormalisation | activation function | dropout | batch size | optimiser | learnrate | pooling | kernel initialiser | deconvolution | DSC |
|---|---|---|---|---|---|---|---|---|---|---|---|---|
| V-Net (4,6,2) | (3, 3, 3) | 16 | True | ELU | 0 | 8 | Adam | $10^{-2}$ | max | Glorot normal | True | 0.92596 |
| V-Net (4,6,2) | (2, 2, 2) | 32 | True | ELU | 0.2 | 8 | Adam | $10^{-3}$ | max | random normal | True | 0.84278 |
| V-Net (4,6,2) | (2, 2, 2) | 32 | True | swish | 0.4 | 32 | RMSProp | $10^{-3}$ | max | random normal | True | 0.07769 |

[a] The model with the highest DSC in each of the architectures tried is highlighted in bold

**Table S11. Blocks, layers and downsampling dimensions hyperparameter tuning for diastole model (two-model segmentation approach)** [a]

| architecture | blocks | layers | dimensions | DSC |
|---|---|---|---|---|
| U-Net++ | 4 | 8 | 3 | 0.92874 |
| U-Net++ | 3 | 4 | 3 | 0.944933333 |
| U-Net++ | 4 | 4 | 2 | 0.920933333 |
| U-Net++ | 3 | 6 | 3 | 0.601446667 |
| **U-Net++** | **3** | **8** | **2** | **0.952003333** |
| U-Net++ | 4 | 6 | 2 | 0.926186667 |
| Attention U-Net | 4 | 4 | 2 | 0.95433 |
| Attention U-Net | 4 | 4 | 3 | 0.95056 |
| Attention U-Net | 4 | 6 | 2 | 0.95357 |
| Attention U-Net | 2 | 4 | 3 | 0.94697 |
| **Attention U-Net** | **4** | **6** | **3** | **0.95544** |
| Attention U-Net | 2 | 8 | 3 | 0.94434 |
| Attention U-Net | 3 | 6 | 2 | 0.95193 |
| Attention U-Net | 3 | 4 | 2 | 0.95018 |
| Attention U-Net | 4 | 2 | 2 | 0.95056 |
| Attention U-Net | 2 | 8 | 3 | 0.86902 |
| V-Net | 2 | 4 | 3 | 0.94244 |
| V-Net | 2 | 6 | 3 | 0.95183 |
| V-Net | 2 | 8 | 3 | 0.94686 |
| V-Net | 2 | 8 | 2 | 0.95652 |
| V-Net | 4 | 8 | 3 | 0.95042 |
| V-Net | 3 | 2 | 3 | 0.94937 |
| **V-Net** | **3** | **4** | **2** | **0.95797** |
| V-Net | 3 | 6 | 3 | 0.95287 |
| V-Net | 2 | 2 | 3 | 0.93696 |
| V-Net | 4 | 6 | 3 | 0.95145 |

[a] The model with the highest DSC in each of the architectures tried is highlighted in bold

**Table S12. Hyperparameter tuning for diastole model (two-model segmentation approach)**[a]

| architecture (blocks, layers, dimensions) | kernel size | number of filters | renormalisation | activation function | dropout | batch size | optimiser | Learning rate | pooling | kernel initialiser | deconvolution | DSC |
|---|---|---|---|---|---|---|---|---|---|---|---|---|
| U-Net++ (3,8,2) | (2, 2, 2) | 32 | FALSE | SELU | 0.4 | 16 | Adam | $10^{-2}$ | max | Glorot normal | TRUE | 0.558886667 |
| U-Net++ (3,8,2) | (3, 3, 3) | 8 | TRUE | GELU | 0.6 | 16 | Adam | $10^{-2}$ | max | Glorot uniform | FALSE | 0.53002 |
| U-Net++ (3,8,2) | (5, 5, 5) | 8 | FALSE | GELU | 0.6 | 32 | RMSProp | $10^{-2}$ | max | Glorot normal | FALSE | 0.594326667 |
| U-Net++ (3,8,2) | (3, 3, 3) | 16 | FALSE | swish | 0.2 | 8 | Adam | $10^{-2}$ | max | Glorot uniform | FALSE | 0.67312 |
| U-Net++ (3,8,2) | (3, 3, 3) | 32 | TRUE | ELU | 0.2 | 32 | Adam | $10^{-3}$ | average | random normal | TRUE | 0.919523333 |
| U-Net++ (3,8,2) | (5, 5, 5) | 8 | TRUE | GELU | 0.2 | 16 | RMSProp | $10^{-2}$ | max | Glorot normal | FALSE | 0.87706 |
| U-Net++ (3,8,2) | (3, 3, 3) | 8 | TRUE | GELU | 0.4 | 32 | RMSProp | $10^{-2}$ | average | random normal | TRUE | 0.61959 |
| U-Net++ (3,8,2) | (5, 5, 5) | 16 | FALSE | GELU | 0.2 | 16 | Adam | $10^{-3}$ | average | random normal | FALSE | 0.714096667 |
| U-Net++ (3,8,2) | (5, 5, 5) | 32 | FALSE | SELU | 0.6 | 8 | Adam | $10^{-3}$ | average | random normal | FALSE | 0.76115 |
| U-Net++ (3,8,2) | (3, 3, 3) | 16 | TRUE | SELU | 0.4 | 32 | Adam | $10^{-2}$ | max | Glorot normal | FALSE | 0.817603333 |
| **U-Net++ (3,8,2)** | **(3, 3, 3)** | **16** | **TRUE** | **ELU** | **0** | **8** | **Adam** | **$10^{-3}$** | **max** | **Glorot normal** | **TRUE** | **0.952003333** |
| U-Net++ (3,8,2) | (3, 3, 3) | 16 | TRUE | swish | 0 | 8 | Adam | $10^{-3}$ | max | Glorot normal | TRUE | 0.89604 |
| U-Net++ (3,8,2) | (5, 5, 5) | 16 | TRUE | ELU | 0.2 | 8 | Adam | $10^{-3}$ | max | Glorot normal | TRUE | 0.65265 |
| U-Net++ (3,8,2) | (3, 3, 3) | 16 | TRUE | ELU | 0 | 8 | Adam | $10^{-3}$ | average | Glorot normal | TRUE | 0.25532 |
| U-Net++ (3,8,2) | (3, 3, 3) | 16 | TRUE | ELU | 0 | 8 | Adam | $10^{-3}$ | max | Glorot normal | FALSE | 0.32035 |
| Attention U-Net (4,6,3) | (2, 2, 2) | 32 | TRUE | swish | 0.2 | 16 | Adam | $10^{-3}$ | average | Glorot normal | FALSE | 0.7399 |

| architecture (blocks, layers, dimensions) | kernel size | number of filters | renormalisation | activation function | dropout | batch size | optimiser | learning rate | pooling | kernel initialiser | deconvolution | DSC |
|---|---|---|---|---|---|---|---|---|---|---|---|---|
| Attention U-Net (4,6,3) | (2, 2, 2) | 8 | TRUE | ELU | 0 | 8 | RMSProp | $10^{-2}$ | average | Glorot uniform | TRUE | 0.72102 |
| Attention U-Net (4,6,3) | (5, 5, 5) | 16 | TRUE | swish | 0.6 | 32 | RMSProp | $10^{-3}$ | max | Glorot uniform | TRUE | 0.14878 |
| Attention U-Net (4,6,3) | (5, 5, 5) | 16 | FALSE | ELU | 0.4 | 32 | RMSProp | $10^{-3}$ | average | Glorot uniform | FALSE | 0.14967 |
| Attention U-Net (4,6,3) | (3, 3, 3) | 32 | FALSE | ELU | 0.6 | 16 | RMSProp | $10^{-2}$ | max | Glorot uniform | FALSE | 0.1461 |
| Attention U-Net (4,6,3) | (3, 3, 3) | 16 | TRUE | ELU | 0 | 8 | Adam | $10^{-3}$ | max | Glorot normal | TRUE | 0.95140 |
| Attention U-Net (4,6,3) | (3, 3, 3) | 16 | TRUE | swish | 0 | 8 | Adam | $10^{-3}$ | max | Glorot normal | TRUE | 0.94496 |
| **Attention U-Net (4,6,3)** | **(3, 3, 3)** | **16** | **TRUE** | **ELU** | **0** | **8** | **Adam** | **$10^{-3}$** | **average** | **Glorot normal** | **TRUE** | **0.95198** |
| Attention U-Net (4,6,3) | (3, 3, 3) | 16 | TRUE | ELU | 0 | 8 | Adam | $10^{-3}$ | max | Glorot normal | FALSE | 0.95007 |
| V-Net (3,4,2) | (5, 5, 5) | 16 | TRUE | ELU | 0.4 | 16 | Adam | $10^{-3}$ | average | Glorot normal | TRUE | 0.9379 |
| **V-Net (3,4,2)** | **(3, 3, 3)** | **16** | **TRUE** | **ELU** | **0** | **8** | **Adam** | **$10^{-3}$** | **max** | **Glorot normal** | **TRUE** | **0.958** |
| V-Net (3,4,2) | (5, 5, 5) | 8 | FALSE | ELU | 0.4 | 16 | Adam | $10^{-3}$ | average | Glorot normal | TRUE | 0.93235 |
| V-Net (3,4,2) | (3, 3, 3) | 32 | TRUE | GELU | 0.4 | 64 | Adam | $10^{-3}$ | max | Glorot uniform | TRUE | 0.19961 |
| V-Net (3,4,2) | (2, 2, 2) | 8 | FALSE | SELU | 0.4 | 32 | Adam | $10^{-3}$ | max | random normal | TRUE | 0.44774 |
| V-Net (3,4,2) | (5, 5, 5) | 16 | FALSE | ELU | 0.6 | 64 | Adam | $10^{-3}$ | max | random normal | TRUE | 0.17005 |
| V-Net (3,4,2) | (2, 2, 2) | 16 | TRUE | SELU | 0.4 | 64 | Adam | $10^{-3}$ | average | random normal | TRUE | 0.82687 |
| V-Net (3,4,2) | (5, 5, 5) | 8 | FALSE | swish | 0.6 | 16 | Adam | $10^{-2}$ | max | random normal | TRUE | 0.89169 |
| V-Net (3,4,2) | (5, 5, 5) | 8 | TRUE | SELU | 0 | 8 | Adam | $10^{-3}$ | max | Glorot uniform | TRUE | 0.92336 |
| V-Net (3,4,2) | (5, 5, 5) | 32 | FALSE | swish | 0.2 | 32 | Adam | $10^{-2}$ | max | Glorot normal | TRUE | 0.39526 |
| V-Net (3,4,2) | (5, 5, 5) | 16 | TRUE | swish | 0.4 | 32 | Adam | $10^{-2}$ | average | Glorot normal | TRUE | 0.93785 |
| V-Net (3,4,2) | (5, 5, 5) | 8 | TRUE | SELU | 0.4 | 32 | Adam | $10^{-3}$ | max | random normal | TRUE | 0.81267 |

*a* The model with the highest DSC in each of the architectures tried is highlighted in bold

**Table S13. Blocks, layers and downsampling dimensions hyperparameter tuning for diastole model (one-model segmentation approach)** [a]

| Architecture | Blocks | Layers | Dimensions | DSC |
|---|---|---|---|---|
| Attention U-Net | 3 | 6 | 3 | 0.86829 |
| Attention U-Net | 2 | 2 | 2 | 0.88033 |
| Attention U-Net | 3 | 8 | 2 | 0.89134 |
| Attention U-Net | 2 | 8 | 2 | 0.90954 |
| Attention U-Net | 2 | 8 | 3 | 0.62085 |
| Attention U-Net | 3 | 2 | 3 | 0.935 |
| Attention U-Net | 3 | 2 | 2 | 0.92557 |
| Attention U-Net | 2 | 8 | 2 | 0.76726 |
| Attention U-Net | 3 | 4 | 2 | 0.92665 |
| V-Net | 4 | 4 | 2 | 0.91871 |
| V-Net | 4 | 2 | 2 | 0.92691 |
| V-Net | 2 | 2 | 2 | 0.92505 |
| V-Net | 3 | 8 | 3 | 0.90504 |
| V-Net | 4 | 8 | 3 | 0.91871 |
| V-Net | 2 | 4 | 2 | 0.88028 |
| V-Net | 4 | 2 | 3 | 0.94203 |
| V-Net | 3 | 2 | 3 | 0.94502 |
| V-Net | 2 | 8 | 3 | 0.9277 |
| U-Net | 4 | 6 | 3 | 0.92226 |
| U-Net | 4 | 4 | 2 | 0.82725 |
| U-Net | 4 | 4 | 2 | 0.9228 |
| U-Net | 2 | 4 | 2 | 0.93748 |
| U-Net | 3 | 8 | 2 | 0.92093 |
| U-Net | 3 | 8 | 2 | 0.93273 |
| U-Net | 3 | 4 | 2 | 0.94318 |
| U-Net | 3 | 6 | 2 | 0.90307 |
| U-Net++ | 2 | 2 | 3 | 0.969726667 |
| U-Net++ | 4 | 4 | 3 | 0.88282 |
| U-Net++ | 3 | 6 | 3 | 0.907096667 |
| U-Net++ | 2 | 4 | 3 | 0.97156 |
| U-Net++ | 4 | 6 | 2 | 0.536096667 |
| U-Net++ | 2 | 2 | 3 | 0.96086 |
| U-Net++ | 3 | 2 | 2 | 0.940423333 |
| U-Net++ | 3 | 6 | 2 | 0.954563333 |
| U-Net++ | 2 | 8 | 2 | 0.974 |
| U-Net++ | 4 | 8 | 2 | 0.92777 |

[a] The model with the highest DSC in each of the architectures tried is highlighted in bold

**Table S14. Hyperparameter tuning for model (one-model segmentation approach)**[a]

| architecture (blocks, layers, dimensions) | kernel size | number of filters | renormalisation | activation function | dropout | batch size | optimiser | learnrate | pooling | kernel initialiser | deconvolution | DSC |
|---|---|---|---|---|---|---|---|---|---|---|---|---|
| U-Net++ (2,8,2) | (5, 5, 5) | 16 | True | ELU | 0.2 | 32 | Adam | $10^{-3}$ | average | Glorot uniform | True | 0.873953 |
| U-Net++ (2,8,2) | (3, 3, 3) | 8 | True | swish | 0.2 | 8 | Adam | $10^{-3}$ | average | Glorot normal | True | 0.774147 |
| U-Net++ (2,8,2) | (3, 3, 3) | 32 | True | GELU | 0 | 8 | Adam | $10^{-3}$ | max | Glorot normal | True | 0.92649 |
| **U-Net++ (2,8,2)** | **(3,3,3)** | **16** | **True** | **ELU** | **0** | **8** | **Adam** | **$10^{-2}$** | **max** | **Glorot normal** | **True** | **0.923** |
| U-Net++ (2,8,2) | (3, 3, 3) | 16 | True | ELU | 0.4 | 32 | Adam | $10^{-3}$ | average | random normal | True | 0.873313 |
| U-Net (3,4,2) | (2, 2, 2) | 32 | True | GELU | 0.2 | 32 | Adam | $10^{-3}$ | average | Glorot uniform | True | 0.55943 |
| U-Net (3,4,2) | (2, 2, 2) | 16 | True | GELU | 0.4 | 32 | Adam | $10^{-2}$ | average | Glorot uniform | True | 0.14882 |
| U-Net (3,4,2) | (3, 3, 3) | 16 | True | SELU | 0.2 | 32 | Adam | $10^{-2}$ | max | Glorot uniform | True | 0.68446 |
| U-Net (3,4,2) | (5, 5, 5) | 32 | True | GELU | 0.4 | 16 | Adam | $10^{-2}$ | average | Glorot uniform | True | 0.63995 |
| **U-Net (3,4,2)** | **(3,3,3)** | **16** | **True** | **ELU** | **0** | **8** | **Adam** | **$10^{-2}$** | **max** | **Glorot uniform** | **True** | **0.94318** |
| U-Net (3,4,2) | (3, 3, 3) | 16 | True | ELU | 0 | 8 | Adam | $10^{-2}$ | max | Glorot normal | True | 0.9269 |
| U-Net (3,4,2) | (5, 5, 5) | 8 | True | GELU | 0.2 | 8 | Adam | $10^{-2}$ | max | random normal | True | 0.91256 |
| V-Net (3,2,3) | (3, 3, 3) | 8 | True | GELU | 0.2 | 8 | Adam | $10^{-2}$ | average | random normal | True | 0.937 |
| V-Net (3,2,3) | (3, 3, 3) | 16 | True | swish | 0 | 32 | Adam | $10^{-3}$ | average | Glorot normal | True | 0.9373 |
| V-Net (3,2,3) | (5, 5, 5) | 8 | True | GELU | 0.4 | 16 | Adam | $10^{-3}$ | max | Glorot normal | True | 0.89056 |
| V-Net (3,2,3) | (3, 3, 3) | 16 | True | GELU | 0.2 | 32 | Adam | $10^{-3}$ | average | Glorot uniform | True | 0.88285 |
| **V-Net (3,2,3)** | **(3,3,3)** | **16** | **True** | **ELU** | **0** | **8** | **Adam** | **$10^{-2}$** | **max** | **Glorot uniform** | **True** | **0.945** |
| V-Net (3,2,3) | (3, 3, 3) | 32 | True | swish | 0 | 8 | Adam | $10^{-2}$ | max | Glorot uniform | True | 0.943 |

*(continuation)*

| architecture (blocks, layers, dimensions) | kernel size | number of filters | renormalisation | activation function | dropout | batch size | optimiser | learnrate | pooling | kernel initialiser | deconvolution | DSC |
|---|---|---|---|---|---|---|---|---|---|---|---|---|
| Attention U-Net (3,2,3) | (3, 3, 3) | 16 | True | GELU | 0.2 | 16 | Adam | $10^{-2}$ | max | Glorot uniform | False | 0.92586 |
| Attention U-Net (3,2,3) | (3, 3, 3) | 32 | True | SELU | 0.4 | 32 | Adam | $10^{-2}$ | max | Glorot normal | False | 0.59272 |
| Attention U-Net (3,2,3) | (3, 3, 3) | 8 | True | GELU | 0.4 | 16 | Adam | $10^{-3}$ | average | random normal | True | 0.37835 |
| **Attention U-Net (3,2,3)** | **(5, 5, 5)** | **16** | **True** | **swish** | **0.2** | **8** | **Adam** | **$10^{-2}$** | **average** | **random normal** | **True** | **0.936** |
| Attention U-Net (3,2,3) | (5, 5, 5) | 32 | True | ELU | 0.2 | 16 | Adam | $10^{-3}$ | average | Glorot uniform | False | 0.85055 |

[a] The model with the highest DSC in each of the architectures tried is highlighted in bold

**Attention U-Net**

The top skip connection in the Attention U-Net both for 1MSA and 2MSA attention coefficients are showed in **Fig. S6**. For both approaches, a trained Attention U-Net is capable of placing more emphasis on the ROI than the milieu comprised of surrounding features, thereby reducing the influence of artefacts and background effects. Nonetheless, it was noticed that they followed different focussing patterns. 1MSA learned to put emphasis on the whole myocardial region both for systole and diastole phases. In contrast, 2MSA only followed this pattern for the diastole phase and for the systole phase only attention coefficients with high values are found in the myocardium contour.

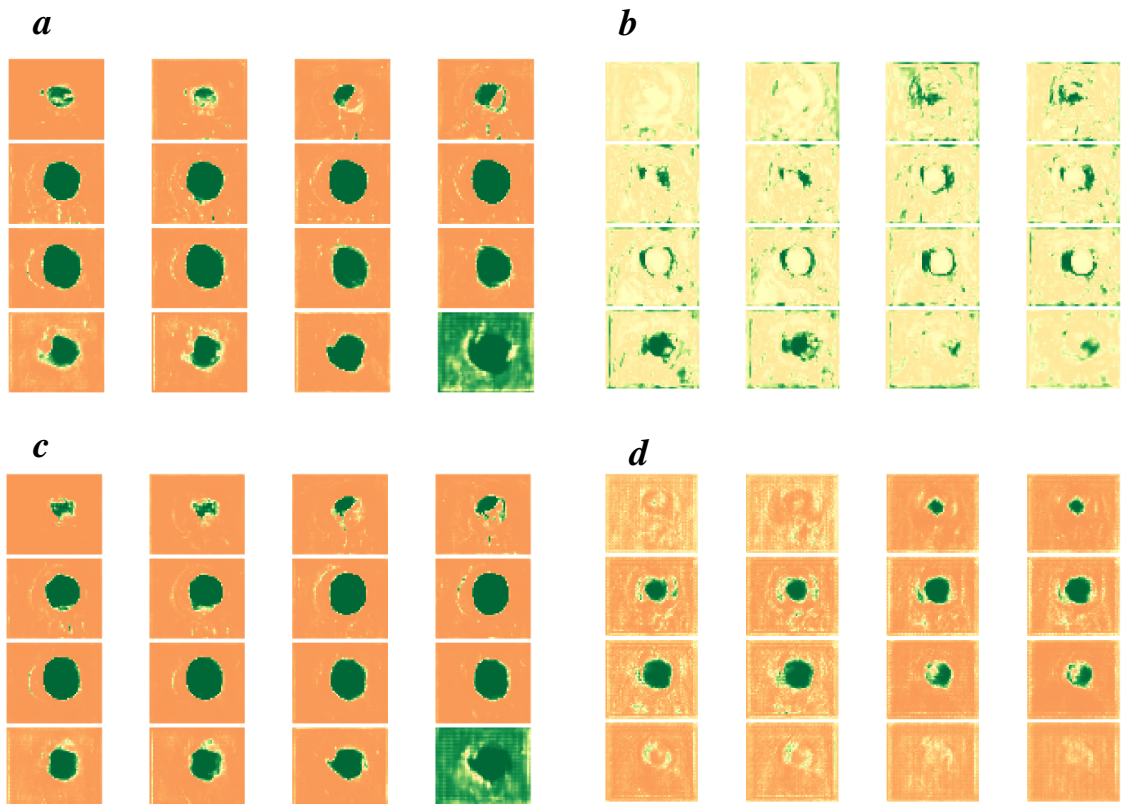

**Figure S6**. Attention coefficient distribution for example images from Study 5.
*a*: 2MSA diastole phase
*b*: 2MSA systole phase
*c*: 1MSA diastole phase
*d*: 1MSA systole phase

# 7. Interoperator Volume Agreement

A close inspection of **Fig. S7**, reveals a strong correlation between the analysed operators. However, it can be easily noticed that Op2 tends to generate segmentations which lead to volume underestimations if compared to Op1. Indeed, this is in line with previous observations made by closely examining the LOESS fit from **Fig. S1**. Interestingly, the diastolic segmentations seem to have a higher degree of agreement than the systolic segmentations.

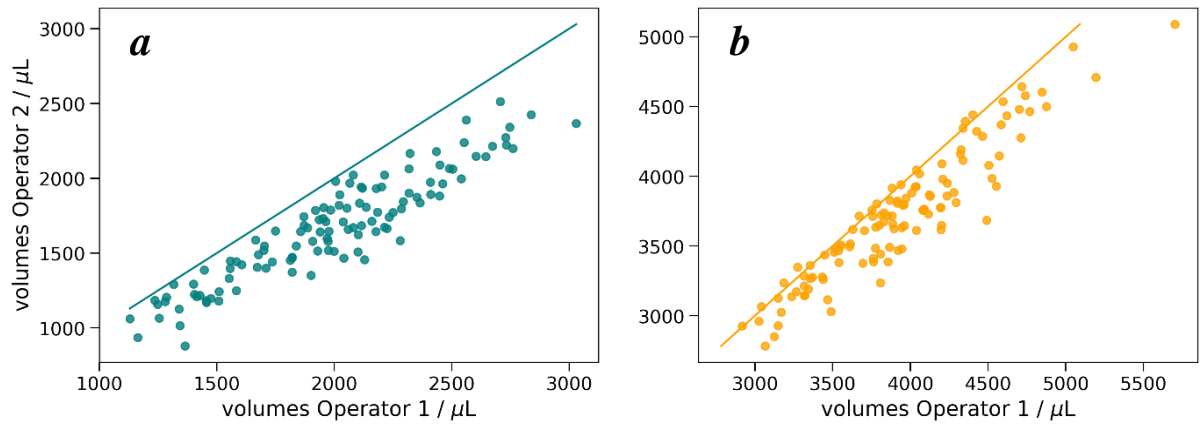

**Figure S7.** Correlation plot between operators. A solid line indicates a perfect correlation.
*a*: Systole volume estimation ($\rho = 0.93$)
*b*: Diastole volume estimation ($\rho = 0.93$)

# 8. Bland-Altman Plots for Automation Feasibility Studies

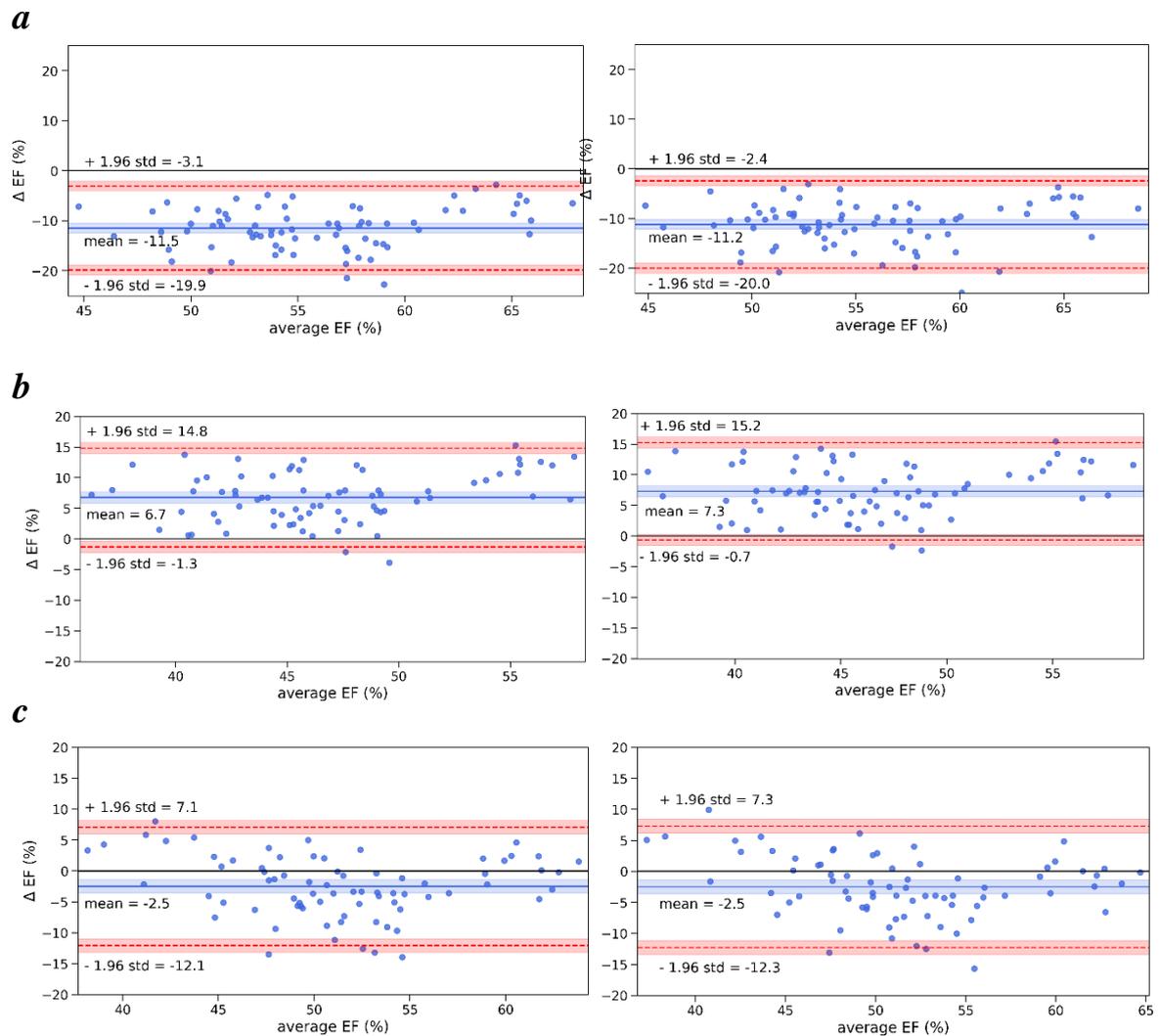

**Figure S8.** Bland-Altman plots for agreement analysis by selection parameter. The bias is marked with a continuous blue line (—) and the equality line is marked with a continuous black line (—). The upper 95% confidence interval bound is marked with a discontinuous green line (--) and the lower one with a discontinuous red line (--). Both are shaded with their respective 95% confidence interval bounds.
***a***: Heart volume as selection parameter. Agreement for polynomial fit displays a bias of -11.5 ± 4.3 % EF (*left*) and agreement for GP fit displays a bias of -11.2 ± 4.3 % EF (*right*).
***b***: Surface area as selection parameter. Agreement for polynomial fit displays a bias of 6.7 ± 4.1 % EF (*left*) and agreement for GP fit displays a bias of 7.4 ± 4.3 % EF (*right*).
***c***: Slice area as selection parameter. Agreement for polynomial fit displays a bias of -2.5 ± 4.9 % EF (*left*) and agreement for GP fit displays a bias of -2.5 ± 5.2 % EF (*right*).